\documentclass[final]{IEEEtran}
\IEEEoverridecommandlockouts

\usepackage{cite}
\usepackage{amsmath,amssymb,amsfonts}

\newcommand{\EqSize}{}          
\newcommand{\CapOpn}{Roots associated to the Cartan matrix}     

\usepackage{algorithmic}
\usepackage{graphicx}
\usepackage{textcomp}
\usepackage{xcolor}

\usepackage{braket,algorithm,amsthm,enumitem}
\usepackage[hidelinks]{hyperref}
\usepackage{tikz}
\usepackage[caption=false,labelformat=simple]{subfig}		

\usetikzlibrary{quantikz}

\newtheorem{definitionenv}{Definition}
\newtheorem{lemmaenv}{Lemma}
\newtheorem{theoremenv}{Theorem}
\newtheorem{corollaryenv}{Corollary}
\newtheorem{propositionenv}{Proposition}
\newtheorem{remarkenv}{Remark}
\newtheorem{conjectureenv}{Conjecture}
\newtheorem{exampleenv}{Example}
\newtheorem{app-lemmaenv}[section]{Lemma}

\newenvironment{definition}{\begin{definitionenv}\rm}{\end{definitionenv}}
\newenvironment{lemma}{\begin{lemmaenv}\rm}{\end{lemmaenv}}

\newenvironment{remark}{\begin{remarkenv}\rm}{\end{remarkenv}}
\newenvironment{example}{\begin{exampleenv}\rm}{\end{exampleenv}}
\newenvironment{proposition}{\begin{propositionenv}\rm}{\end{propositionenv}}

\newenvironment{app-lemma}{\begin{app-lemmaenv}\rm}{\end{app-lemmaenv}}

\DeclareMathOperator*{\argmin}{arg\,min}
\DeclareMathOperator{\wt}{wt}

\newcommand{\hm}{{\varphi}}
\newcommand{\hmi}{{\varphi^{-1}}}

\newcommand{\bNull}{ }
\newcommand{\bZero}{\mathbf{0}}

\newcommand{\bA}{ \mathbf{A}}

\newcommand{\bE}{ \mathbf{E}}
\newcommand{\bF}{ \mathbf{F}}
\newcommand{\bU}{ \mathbf{U}}

\newcommand{\ba}{ \mathbf{a}}

\newcommand{\be}{ \mathbf{e}}
\newcommand{\bff}{ \mathbf{f}}
\newcommand{\bg}{ \mathbf{g}}
\newcommand{\bs}{ \mathbf{s}}

\newcommand{\bv}{ \mathbf{v}}

\newcommand{\Pn}{ \{I,X,Y,Z\}^n}

\newcommand{\sM}{{\cal M}}
\newcommand{\sN}{{\cal N}}
\newcommand{\sS}{{\cal S}}

\newcommand{\cD}{{\cal D}}

\newcommand{\cG}{{\cal G}}

\newcommand{\cS}{{\cal S}}
\newcommand{\cC}{{\cal C}}

\newcommand{\mC}{{\mathbb C}}
\newcommand{\mR}{{\mathbb R}}
\usepackage{accents} 		

\makeatletter 
\renewcommand*\env@matrix[1][*\c@MaxMatrixCols c]{%
  \hskip -\arraycolsep
  \let\@ifnextchar\new@ifnextchar
  \array{#1}}
\makeatother

\usepackage[colorinlistoftodos,prependcaption]{todonotes}
\usepackage{xcolor}
\usepackage{xargs}

\newcommandx{\rednote}[2][1=]{\todo[inline,linecolor=red,backgroundcolor=red!25,bordercolor=red,#1]{#2}}
\newcommandx{\yellownote}[2][1=]{\todo[inline,linecolor=yellow,backgroundcolor=yellow!25,bordercolor=yellow,#1]{#2}}

\def\BibTeX{{\rm B\kern-.05em{\sc i\kern-.025em b}\kern-.08em
    T\kern-.1667em\lower.7ex\hbox{E}\kern-.125emX}}
\begin{document}


\title{Generalized quantum data-syndrome codes and belief propagation decoding for phenomenological noise}

\author{
Kao-Yueh Kuo \,and\, Ching-Yi Lai
\thanks{\footnotesize
This article was presented in part at the 2024 IEEE International Sympoium on Information Theory (ISIT)~\cite{KL24isit}

Kao-Yueh Kuo acknowledges support from Engineering and Physical Sciences Research Council (EPSRC) (Grant No. EP/W028115/1).
Ching-Yi Lai was supported by the National Science and Technology Council in Taiwan under Grants Nos. 111-2628-E-A49-024-MY2, 112-2119-M-A49-007,  112-2119-M-001-007, 113-2221-E-A49-114-MY3 and 113-2119-M-A49-008.

Kao-Yueh Kuo is with the School of Mathematical and Physical Sciences, University of Sheffield, Sheffield S3 7RH, UK. (email: kywukuo@gmail.com)

Ching-Yi Lai is with the Institute of Communications Engineering, National Yang Ming Chiao Tung University, Hsinchu 30010, Taiwan.   (email: cylai@nycu.edu.tw) 
}
}
\maketitle

\thispagestyle{plain}
\pagestyle{plain}

\begin{abstract} 
Quantum stabilizer codes often struggle with syndrome errors due to measurement imperfections. Typically, multiple rounds of syndrome extraction are employed to ensure reliable error information. In this paper, we consider phenomenological decoding problems, where data qubit errors may occur between extractions, and each   measurement can be faulty. 
We introduce generalized quantum data-syndrome  codes along with a generalized check matrix that integrates both quaternary and binary alphabets to represent diverse error sources. This results in  a Tanner graph with mixed variable nodes,  enabling the design of belief propagation (BP) decoding algorithms that effectively handle phenomenological errors. Importantly, our BP decoders are applicable to general sparse quantum codes.
Through simulations, we achieve an error threshold of more than 3\% for quantum memory protected by rotated toric codes, using solely BP without post-processing. Our results indicate that $d$ rounds of syndrome extraction are sufficient for a toric code of distance $d$. We observe that at high error rates, fewer rounds of syndrome extraction tend to perform better, while more rounds improve performance at lower error rates.
Additionally, we propose a method to construct effective redundant stabilizer checks for single-shot error correction. Our simulations show that BP decoding remains highly effective even with a high syndrome error rate.

\end{abstract}


\section{Introduction}

Quantum information has found applications beyond classical systems \cite{BB84,BW92,BBCJPW93, Shor94,Gro96, Cir+97,Bri+98, FGG14,Per+14}. Nevertheless, quantum states are inherently fragile and require protection through quantum error correction (QEC)~\cite{Shor96,GotPhD,CRSS98,NC00}. Encoding quantum states into a larger quantum space, defined by a stabilizer group, enables the detection and correction of deviations from the allowed logical states due to quantum noise.  The stabilizers are then measured and their results are used to deduce the errors that have occurred \cite{DKLP02,MMM04}. These outcomes, known as error syndromes, are pivotal in the QEC process.  However, quantum operations are imperfect, which presents a challenge in achieving practical quantum computation in current techniques~\cite{Aru+19,CFYW19, Rya+21,Abo+22,Kri+22, Kim+23}. Specifically, quantum measurement errors appear to be the predominant source of errors~\cite{Aru+19,CFYW19}. This adds complexity to QEC because it necessitates the syndrome extraction   through imperfect measurements in order to perform effective recovery operations.

The QEC coding theory has been expanded to include scenarios involving both data qubit and syndrome extraction errors~\cite{ALB14,ALB16,ALB20,Fuj14a, ZAWP19, GGKL21}. 
In the framework of quantum data-syndrome (DS) codes~\cite{ALB20}, it is assumed that, in addition to errors affecting the data qubits initially, error syndromes can also be subject to measurement errors. 
To address this, redundant syndrome measurements are introduced to diagnose measurement errors. 
This is also known as single-shot error correction \cite{Bom15,Cam19, QVRC21,BL22, HB23, KV22, DRS22}.

A more realistic error model incorporates the occurrence of data errors between two syndrome extractions. This model, referred to as the phenomenological noise model~\cite{DKLP02,WHP03,Step14s}, goes beyond traditional quantum coding theory. It provides an abstract representation of errors, simplifying the complex circuit-level noise model, where every location in a quantum circuit could potentially be faulty. Despite its abstraction, the phenomenological noise model effectively encapsulates the fundamental challenge of fault-tolerant error correction in the context of circuit-level noise.

We aim to explore the lifetime of a quantum memory in the phenomenological noise model protected by QEC. Relevant topics related to this error model have been investigated in the literature~\cite{ZAWP19,GGKL21,Bom15,Cam19,QVRC21,BL22,HB23,KV22,DRS22}. However, two critical issues persist in most literature.
Firstly, while additional encoded syndrome measurements enhance BP decoding performance, their usage should be restricted due to architectural constraints. For instance, obtaining the syndrome measurement of a nonlocal stabilizer poses significant challenges.
Secondly, a quantum memory is designed for a prolonged lifetime that may involve over $10^{10}$ rounds of syndrome measurements. Thus, buffering all syndrome history for single-shot error correction  becomes impractical \cite{DKLP02}. Single-shot correction can only verify a provable threshold \cite{Ter15,BNB16}. For accurate estimation of the error threshold in practical scenarios, periodic cycles of corrections should be performed, and the quantum memory is deemed viable as long as its residual errors remain correctable.

In this paper, we extend the framework of DS codes to accommodate the phenomenological noise model. 
The  error correction capabilities of a quantum code primarily rely on the check matrix.
However, when multiple rounds of syndrome extraction are considered,   data errors are introduced before each round of syndrome extraction, and the syndrome outcomes themselves can be noisy.
To address these diverse sources of errors, we introduce a generalized check matrix that operates over mixed quaternary and binary alphabets.
This matrix serves to represent the error syndromes associated with each of these error sources.
Subsequently, we formulate decoding problems for the phenomenological noise model using this generalized check matrix.
This allows us to employ established coding techniques for problem resolution.
Specifically, the generalized check matrix gives rise to a Tanner graph comprising quaternary and binary variable nodes, along with conventional binary check nodes.

We propose the use of belief propagation (BP) decoders to address the challenges posed by phenomenological decoding problems.
Gallager's sum-product algorithm, also understood as Pearl's BP, has proven to be effective for classical low-density parity-check (LDPC) codes~\cite{Gal63,MN96,Mac99,Pea88}.
It~is highly efficient with complexity almost linear in the code length.
BP operates as an iterative message-passing algorithm running on the Tanner graph induced by the check matrix of a block code~\cite{Tan81,Pea88,KFL01}. 
It has been extended to decode quantum stabilizer codes as well~\cite{MMM04,PC08,Wan+12,Bab+15, PK19,RWBC20, KL20,KL21a,KL21,KL22isit,yao2024belief,chytas2024enhanced}.
Additionally, a DS-BP algorithm has been proposed to handle quantum DS codes~\cite{KCL21}, where a Tanner graph with mixed alphabets is constructed, and the implementation of BP for unified data-syndrome decoding has been explored.

For phenomenological quantum memory using degenerate quantum codes, DS-BP is ineffective due to the numerous short cycles in the generalized check matrix. 
Recently, refined BP with additional memory effects (called MBP) has shown excellent performance even for degenerate quantum codes~\cite{KL21,KL22isit}. 
Extensive testing across various stabilizer code families has demonstrated that MBP and its adaptive version, AMBP, offer robust decoding and reliable threshold performance, with reduced computational complexity relative to alternative methods. Building on this, we extend MBP and AMBP to generalized DS-MBP (GDS-MBP) and generalized DS-AMBP (GDS-AMBP), respectively, for decoding under the phenomenological noise model.
Like DS-BP, GDS-(A)MBP performs joint decoding of both quaternary data qubit errors and binary syndrome flip errors, using only scalar messages.

Our proposed decoding schemes are suitable for both single-shot and multiple-round QEC scenarios. Notably, our decoding methods do not require post-processing techniques such as ordered statistics decoding \cite{PK19,RWBC20,QVRC21,BL22,HB23} or small-set-flip \cite{GGKL21}. Additionally, the unified data-syndrome decoding process eliminates the need for a two-step decoding \cite{ZAWP19,KV22}. Moreover, our decoding techniques are not confined to topological codes \cite{BNB16,BDMT16,KP19}, as they are applicable to general codes.

In our investigation of the lifetime of a quantum memory protected by quantum stabilizer codes in the phenomenological noise model, general quantum codes based on sparse graphs \cite{MMM04,TZ09_14,KP13,BH14,LTZ15,PK19,PK20} are considered as candidates, utilizing GDS-(A)MBP decoding. However, we specifically account for practical physical limitations, such as 2D layouts or short-range interactions. Consequently, we focus  on quantum memory protected by 2D topological codes \cite{Kit97_03,DKLP02,BM06,BM07,HFDM12}, including rotated toric codes.
Through computer simulations using GDS-AMBP on rotated toric codes, we observe an error threshold of more than 3\% in the phenomenological noise model.
Here, we assume that measurement outcomes are independently flipped with the same rate as the independent depolarizing error rate on the data qubits.
 GDS-AMBP offers a complexity that is almost linear in the code length in addition to a good error threshold.

In the context of single-shot  QEC, it is possible to  achieve excellent performance with only a single round of noisy syndrome extraction   if redundant checks can be flexibly designed \cite{ALB14,ALB16,ALB20}. To align with realistic scenarios, it is essential to have a sparse DS check matrix with few redundant checks.  Previous work~\cite{Fuj14a,KCL21}  has focused on this scenario and demonstrated that, for distance-3 codes, effective performance can be maintained even in the presence of syndrome errors without introducing excessive redundancy. 
Distance-3 codes can correct one syndrome error as long as the columns of the parity check matrix have weights higher than one. However, for higher distance codes with redundant parity checks, it is not obvious whether DS-BP works, as the parity check matrix would have more induced short cycles.

In this paper, we propose a construction that introduces a limited number of redundant checks using random quasi-cyclic matrices.  
To  demonstrate the effectiveness of GDS-(A)MBP on our designed check matrices, we simulate a generalized bicycle (GB) code with a distance of eight for single-shot QEC. When the syndromes are perfect, our numerical results indicate that (A)MBP can correct most errors of weight up to 12, even though that the code  guarantees to correct  errors of weight up to three. 
When the syndrome and data error rates are equal, we observe that most errors of weight  up to 12, encompassing both data and syndrome errors, can be corrected in a single shot. Furthermore, our decoding performance is robust, as most errors of weight  up to four can be corrected, even when the syndrome error rate is ten times the data error rate.

The paper is structured as follows:
In Section~\ref{sec:basics}, we provide an introduction to quantum stabilizer codes and quantum DS codes.
Section~\ref{sec:gen_DS} is dedicated to the definition of phenomenological decoding problems and the proposal for estimating the lifetime of a quantum memory.
In Section~\ref{sec:dec}, we provide the GDS-(A)MBP decoding algorithms.
The complexity analysis of GDS-(A)MBP and the simulation results for quantum memory in rotated toric codes are presented in Section~\ref{sec:mem_QEC}.
Section~\ref{sec:Single_QEC} covers the design criteria and numerical results for single-shot QEC of a GB code.
The conclusion can be found in Section~\ref{sec:con}.

\section{Preliminaries} \label{sec:basics}

\subsection{Basics of quantum stabilizer codes}

Let $I=[{1\atop 0}{0\atop 1}],\, X=[{0\atop 1}{1\atop 0}],\, Y=[{0\atop i}{-i\atop 0}], \text{ and } Z=[{1\atop 0}{0\atop -1}]$ be the Pauli matrices.
The $n$-fold Pauli group under multiplication is
	$
	\cG_n = \{ c\, E_1 \otimes \dots \otimes E_n: c\in\{\pm1\pm i\}, E_j\in\{I,X,Y,Z\} \}.
	$ 
 The weight of an operator $E\in\cG_n$, denoted  $\wt(E)$, is  the number of Pauli components of $E$ that are non-identity. 
We use $X_j$ or $Z_j$ to represent a nontrivial Pauli matrix applied to qubit $j$,
while operating trivially  on the other qubits.
We will only consider quantum errors that are Pauli operators.

Any two Pauli operators either commute or anticommute. 
Let $\cS$ be an abelian subgroup of $\cG_n$ such that $-I^{\otimes n}\notin \cS$.
Suppose that $\cS$ is generated by $n-k$ independent generators.
Then the joint $(+1)$-eigenspace of~$\cS$ 
	$$
	\cC(\cS) = \{\ket{\psi}\in\mC^{2^n}: g\ket{\psi} = \ket{\psi} ~\forall\, g\in\cS\}
	$$
is called an $[[n,k]]$ stabilizer code, which encodes $k$ logical qubits into $n$ physical qubits.
The group $\cS$ is called a stabilizer group and its elements  are called \textit{stabilizers} for $\cC(\cS)$.
The stabilizers can be measured on a code state without perturbing its quantum information. A Pauli error $E\in\cG_n$   can be detected by $\cC(\cS)$  if $E$ anticommutes with any of the stabilizers in $\cS$.  
Let
	$$
	{ \cS^\perp= \{E\in\cG_n: gE=Eg ~\forall\, g\in\cS\} }
	$$
denote the set of elements in $\cG_n$ that commute with $\cS$. 
As the errors in $\cS^\perp$ cannot be detected and may change  logical states, 
errors in $\cS^\perp\setminus (\{\pm 1, \pm i\}\times\cS)$ are called \textit{logical errors}
and the minimum distance of   $\cC(\cS)$ is
	$$
	d = \min\{\wt(E): E\in\cS^\perp\setminus (\{\pm 1, \pm i\}\times\cS) \}.
	$$
$\cC(\cS)$ is called an $[[n,k,d]]$ stabilizer code.
It is understood that any error $E\in\cG_n$ of weight $\wt(E)\le \lfloor \frac{d-1}{2} \rfloor$ can be corrected by the code~\cite{KL97}.

Because  our examination of decoding problems is independent of the phase of a Pauli operator,
we define a homomorphism $\hm: \cG_n\rightarrow \{I,X,Y,Z\}^n$ as follows:   
for an $n$-fold Pauli operator $ E=c E_1 \otimes  \dots \otimes E_n\in\cG_n$,
$$\hm(E)=(E_1, \dots,E_n) \in\{I,X,Y,Z\}^n$$ is a  row vector, regardless of the phase in $E$. 
For $\bE=(E_1,\dots, E_n)\in\{I,X,Y,Z\}^n$, define $$\hm^{-1}(\bE)= E_1\otimes \cdots \otimes E_n.$$
For $\bE=(E_1,\dots, E_n), \bF=(F_1,\dots, F_n)\in\{I,X,Y,Z\}^n$, the product of $\bE$ and $\bF$
should be understood as 
\begin{align}
\bE\bF= ( \hm(E_1 F_1),\dots, \hm(E_n F_n))\in\{I,X,Y,Z\}^n.
\end{align}
 Throughout the context,  we will almost  work with  vectors in $\Pn$.

We define a bilinear form on $\{I,X,Y,Z\}^n$ as follows: for $\bE,\bF\in \{I,X,Y,Z\}^n$,
\begin{align}
\bE*\bF=	
\begin{cases} 
	0, &\text{if $\hmi(\bE)$ and $\hmi(\bF)$ commute},\\
	1, &\text{if $\hmi(\bE)$ and $\hmi(\bF)$ anticommute}.
\end{cases}
\end{align}

Assume that a Pauli error $\bE\in\Pn$ occurs on a codeword of  $\cC(\cS)$.
Suppose that $m\ge n-k$ stabilizers with corresponding vectors $\bg_1,\bg_2,\dots,\bg_m\in\Pn$ are perfectly measured and  the binary outcomes are $\bs=(s_1,\dots,s_m)\in\{0,1\}^m$, where
	$$
	s_i = \bE*\bg_i. 
	$$
The binary vector $\bs$ is called the error syndrome of $E$.

Assume that $\bg_i = (H_{i1}, H_{i2}, \dots, H_{in})\in\{I,X,Y,Z\}^n$. 
A check matrix for the stabilizer code $\cC(\cS)$ is
	\begin{align}
	  {H} = [H_{ij}]
	\in\{I,X,Y,Z\}^{m\times n} \label{eq:S}
	\end{align}
 so the $i$-th row of $H$ is $\bg_i$. 
For simplicity, let $\bs =  \bE*H \in \{0,1\}^m$ denote that $s_i =  \bE* \bg_i$ for $i=1,\dots,m$.

If two errors $\bE$ and $\bE'\in\Pn$   are equal up to a stabilizer, they are equivalent on the codespace  and thus  are referred to as  \textit{degenerate} errors of each other. 
The presence of this degeneracy should be considered when addressing the decoding problem for a stabilizer code.

\begin{definition}{\bf (Stabilizer decoding problem)} \label{def:dec}
{\bf Stabilizer decoding problem:} 
Given a check matrix $ {H}\in\{I,X,Y,Z\}^{m\times n}$ and a syndrome $\bs\in\{0,1\}^m$ of an error $\bE\in\Pn$, output  $\hat \bE\in \Pn$ such that $  \hat \bE* {H} = \bs$ and $ \hat \bE \bE\in \hm(\sS)$.
\qed
\end{definition}

This general decoding problem is known to be NP-hard \cite{KL13_20}, even \#P-hard \cite{IP15}. However, for specific sparse codes, there are efficient decoders, such as MWPM decoder designed for topological toric or surface codes \cite{DKLP02,WHP03}, as well as BP decoders for general sparse-graph codes~\cite{PC08,KL21}.

\subsection{Basics of quantum data-syndrome codes}

In the context of quantum DS codes \cite{ALB14,ALB16,ALB20}, the error model under consideration involves both data qubit errors and syndrome measurement errors.

Now, let us consider that in addition to an $n$-fold depolarizing Pauli error $\bE\in\Pn$ occurring on the data qubits of a codeword, during the syndrome extraction of the $m$ stabilizers $\{\bg_i\}_{i=1}^{m}$, each syndrome bit $s_i$ is subject to an independent bit-flip error $e_i\in\{0,1\}$.
Thus we have error syndrome 
	\begin{equation} \label{eq:nos_s_i}
	s_i =    \bE*\bg_i + e_i \mod 2, \quad \text{for $i=1,2,\dots,m$.}
	\end{equation}

This process can be expressed in the context of a code over joint alphabets $\{I,X,Y,Z\}$ and $\{0,1\}$.
Define a bilinear form for  two vectors $(\bE,\be),(\bF,\bff)\in \{I,X,Y,Z\}^n\times \{0,1\}^m$  by 
\begin{align*} 
	 (\bE,\be)*(\bF,\bff)=  \bE*\bF+\be\cdot \bff   \mod 2,
\end{align*}
where 	$\textstyle \be\cdot \bff = \sum_{i=1}^m e_if_i \mod 2$ for  $\be,\bff\in\{0,1\}^m$ 
and we reuse the notation $*$ without ambiguity. 
Then   a DS check matrix is defined as  
	\begin{equation} \label{eq:DS_check}
	\tilde{ {H}}  = [ {H} \,|\,  {I}_m]\in\{I,X,Y,Z\}^{m\times n}\times \{0,1\}^{m\times m},
	\end{equation}
where $ {I}_m$ is an $m\times m$ identity matrix.
We say that $\tilde{H}$ defines a quantum DS code, which characterizes the algebraic structure of a quantum code and its syndrome relations.  
The relation between the error vector $(\bE,\be)$ and its error syndrome $\bs$ in \eqref{eq:nos_s_i} can then be written as 
$$
s_i =   (\bE,\be)* \tilde \bg_i,
$$
where  $\tilde{\bg}_i$ is the $i$-th row of $\tilde{ {H}}$.
Similarly, let $\bs =  (\bE,\be)* \tilde{ {H}}  $ denote that $s_i =   (\bE,\be)*\tilde \bg_i  $ for all~$i$.

\begin{definition} {\bf(DS decoding problem)} \label{def:dsdp}
Given a DS check matrix $\tilde{ {H}} = [ {H}\,|\, {I}_m]$, where $ {H}\in\{I,X,Y,Z\}^{m\times n}$, and a syndrome $\bs\in\{0,1\}^m$ of an error $(\bE,\be)\in\{I,X,Y,Z\}^n\times \{0,1\}^m$,  output  $(\hat \bE, \hat \be)\in \{I,X,Y,Z\}^n \times \{0,1\}^m$ such that $ (\hat \bE,\hat \be) * \tilde{ {H}}  = \bs$ and $\hat \bE \bE\in \hm(\sS)$.
\qed
\end{definition}

It is evident that the error correction capabilities  {on the data qubits} of a quantum DS code are affected by the presence of syndrome errors. However, it is possible to maintain the same level of error correction guarantee on the data qubits by introducing additional syndrome measurements, as discussed in \cite{ALB16,ALB20}.

A DS check matrix gives rise to a Tanner graph, which can be used for iterative decoding, as discussed in \cite{KCL21}. For an $n$-qubit stabilizer code with   $m$ stabilizers being measured, the Tanner graph consists of $m+n$ variable nodes, corresponding to an error vector $(\bE,\be)\in\{I,X,Y,Z\}^{n}\times\{0,1\}^{m}$, and $m$ check nodes, corresponding a syndrome vector $\bs\in\{0,1\}^m$. Edges $(i,j)$ connect variable node $j$ with check node $i$ if either the $(i,j)$-th entry of $\tilde{ {H}}$, denoted as $\tilde H_{ij}$,  {satisfies} $\tilde H_{ij}\in \{X,Y,Z\}$ for $j\le n$ or $\tilde H_{ij}=1$ for $j>n$.

We remark that a quantum DS code can be equivalently defined by various check matrices. However, the induced Tanner graphs will differ, and some of these graphs may be better suited for decoding purposes.

\begin{example} \label{ex:DS_exs}
Consider a check matrix 
	$
	\left[\begin{smallmatrix}
	X & Y & I \\ Z & Z & Y
	\end{smallmatrix}\right]
	$,
 which induces a DS check matrix 
	$
	\left[\begin{smallmatrix}
	X & Y & I \\ Z & Z & Y
	\end{smallmatrix}\middle|
	\begin{smallmatrix}
	1 & 0 \\ 0 & 1
	\end{smallmatrix}\right]
	$.
To enhance syndrome protection, we  introduce one additional redundant stabilizer 
 $(Y, X,  Y)$, resulting a modified DS check matrix 
	$\tilde{ {H}}= 
	\left[\begin{smallmatrix}
	X & Y & I \\ Z & Z & Y\\ Y & X & Y
	\end{smallmatrix}\middle|
	\begin{smallmatrix}
	1 & 0 & 0 \\ 0 & 1 & 0 \\ 0 & 0 & 1
	\end{smallmatrix}\right]
	$.
By applying certain row operations to $\tilde H$, we obtain an alternative DS check matrix
	$\tilde{ {H}}'= 
	\left[\begin{smallmatrix}
	X & Y & I \\ Z & Z & Y\\ I & I & I
	\end{smallmatrix}\middle|
	\begin{smallmatrix}
	1 & 0 & 0 \\ 0 & 1 & 0 \\ 1 & 1 & 1
	\end{smallmatrix}\right].
	$
        
 The corresponding Tanner graphs of the DS matrices  are depicted in Figure~\ref{fig:DS_exs}.
Both DS check matrices $\tilde{ {H}}$ and $\tilde{ {H}}'$ define the same DS code.
However, the third check in $\tilde{ {H}}'$ signifies that  {the modulo-2 sum of the three syndrome errors should equal the third syndrome $s'_3$ of $\tilde{ {H}}'$}.
This additional information may prove helpful in the decoding process.
\qed
\end{example}

\begin{figure}
	\centering
	\resizebox{0.48\textwidth}{!}{ 
		\subfloat[\label{fig:DS_ex1}]{\begin{tikzpicture}[node distance=1.3cm,>=stealth,bend angle=45,auto]

\tikzstyle{chk}=[rectangle,thick,draw=black!75,fill=black!20,minimum size=4mm]
\tikzstyle{var}=[circle,thick,draw=blue!75,fill=gray!20,minimum size=4mm,font=\footnotesize]
\tikzstyle{VAR}=[circle,thick,draw=blue!75,fill=blue!20,minimum size=5mm,font=\footnotesize]
\tikzstyle{fac}=[anchor=west,font=\footnotesize]

\node[var] (x1) at (0,2) {$E_1$};
\node[var] (x2) at (1,2) {$E_2$};
\node[var] (x3) at (2,2) {$E_3$};
\node[chk] (c1) at (0,0.9) {};
\node[chk] (c2) at (1,0.9) {};
\node[var] (z1) at (0,0) {$e_1$};
\node[var] (z2) at (1,0) {$e_2$};

\draw[thick] (x1) -- (c1);
\draw[thick,dashed] (x2) -- (c1);
\draw[thick,densely dotted] (x1) -- (c2);
\draw[thick,densely dotted] (x2) -- (c2);
\draw[thick,dashed] (x3) -- (c2);
\draw[thin] (c1) -- (z1);
\draw[thin] (c2) -- (z2);

\node[fac] (Xl) [right of=c2,xshift= 0-20,yshift=  0+8] {};
\node[fac] (Xr) [right of=c2,xshift=26-20,yshift=  0+8] {$X$};
\node[fac] (Yl) [right of=c2,xshift= 0-20,yshift= -6+8] {};
\node[fac] (Yr) [right of=c2,xshift=26-20,yshift= -6+8] {$Y$};
\node[fac] (Zl) [right of=c2,xshift= 0-20,yshift=-12+8] {};
\node[fac] (Zr) [right of=c2,xshift=26-20,yshift=-12+8] {$Z$};
\draw[thick] (Xl) -- (Xr);
\draw[thick,dashed] (Yl) -- (Yr);
\draw[thick,densely dotted] (Zl) -- (Zr);

\end{tikzpicture}}\quad~
		\subfloat[\label{fig:DS_ex2}]{\begin{tikzpicture}[node distance=1.3cm,>=stealth,bend angle=45,auto]

\tikzstyle{chk}=[rectangle,thick,draw=black!75,fill=black!20,minimum size=4mm]
\tikzstyle{var}=[circle,thick,draw=blue!75,fill=gray!20,minimum size=4mm,font=\footnotesize]
\tikzstyle{VAR}=[circle,thick,draw=blue!75,fill=blue!20,minimum size=5mm,font=\footnotesize]
\tikzstyle{fac}=[anchor=west,font=\footnotesize]

\node[var] (x1) at (0,2) {$E_1$};
\node[var] (x2) at (1,2) {$E_2$};
\node[var] (x3) at (2,2) {$E_3$};
\node[chk] (c1) at (0,0.9) {};
\node[chk] (c2) at (1,0.9) {};
\node[chk] (c3) at (2,0.9) {};
\node[var] (z1) at (0,0) {$e_1$};
\node[var] (z2) at (1,0) {$e_2$};
\node[var] (z3) at (2,0) {$e_3$};

\draw[thick] (x1) -- (c1);
\draw[thick,dashed] (x2) -- (c1);
\draw[thick,densely dotted] (x1) -- (c2);
\draw[thick,densely dotted] (x2) -- (c2);
\draw[thick,dashed] (x3) -- (c2);
\draw[thick,dashed] (x1) -- (c3);
\draw[thick] (x2) -- (c3);
\draw[thick,dashed] (x3) -- (c3);
\draw[thin] (c1) -- (z1);
\draw[thin] (c2) -- (z2);
\draw[thin] (c3) -- (z3);

\end{tikzpicture}}\qquad
		\subfloat[\label{fig:DS_ex3}]{\begin{tikzpicture}[node distance=1.3cm,>=stealth,bend angle=45,auto]

\tikzstyle{chk}=[rectangle,thick,draw=black!75,fill=black!20,minimum size=4mm]
\tikzstyle{var}=[circle,thick,draw=blue!75,fill=gray!20,minimum size=4mm,font=\footnotesize]
\tikzstyle{VAR}=[circle,thick,draw=blue!75,fill=blue!20,minimum size=5mm,font=\footnotesize]
\tikzstyle{fac}=[anchor=west,font=\footnotesize]

\node[var] (x1) at (0,2) {$E_1$};
\node[var] (x2) at (1,2) {$E_2$};
\node[var] (x3) at (2,2) {$E_3$};
\node[chk] (c1) at (0,0.9) {};
\node[chk] (c2) at (1,0.9) {};
\node[chk] (c3) at (2,0.9) {};
\node[var] (z1) at (0,0) {$e_1$};
\node[var] (z2) at (1,0) {$e_2$};
\node[var] (z3) at (2,0) {$e_3$};

\draw[thick] (x1) -- (c1);
\draw[thick,dashed] (x2) -- (c1);
\draw[thick,densely dotted] (x1) -- (c2);
\draw[thick,densely dotted] (x2) -- (c2);
\draw[thick,dashed] (x3) -- (c2);
\draw[thin] (c1) -- (z1);
\draw[thin] (c2) -- (z2);
\draw[thin] (c3) -- (z1);
\draw[thin] (c3) -- (z2);
\draw[thin] (c3) -- (z3);

\end{tikzpicture}}
	}
        \caption[\CapOpn]{
		The Tanner graphs corresponding to
		(a)~${
			\left[\begin{smallmatrix}
			X & Y & I \\ Z & Z & Y
			\end{smallmatrix}\middle|
			\begin{smallmatrix}
			1 & 0 \\ 0 & 1
			\end{smallmatrix}\right],
			}$
		(b)~${ 
			\left[\begin{smallmatrix}
			X & Y & I \\ Z & Z & Y \\ Y & X & Y
			\end{smallmatrix}\middle|
			\begin{smallmatrix}
			1 & 0 & 0 \\ 0 & 1 & 0 \\ 0 & 0 & 1
			\end{smallmatrix}\right],
			}$
		and
		(c)~${ 
			\left[\begin{smallmatrix}
			X & Y & I \\ Z & Z & Y \\ I & I & I
			\end{smallmatrix}\middle|
			\begin{smallmatrix}
			1 & 0 & 0 \\ 0 & 1 & 0 \\ 1 & 1 & 1
			\end{smallmatrix}\right]
			}.$
   Variable nodes and check nodes are represented as circles and boxes, respectively.
	} \label{fig:DS_exs}
\end{figure}

For DS codes of small distance, it is possible to design a BP decoder for DS codes, which can achieve strong performance without the need for numerous redundant syndrome measurements~\cite{KCL21}.

\section{Phenomenological noise model}  \label{sec:gen_DS}

In a more practical error model, especially when a substantial number of stabilizer measurements are conducted, there is a risk of data errors accumulating, especially if the quantum memory error rate is high. When a specific set of stabilizers is repeated multiple times to obtain reliable error syndromes, a phenomenological noise model assumes  that data errors occur between rounds of measuring the same set of stabilizers, in addition to measurement errors.

Consider a quantum codeword from an $[[n,k]]$ stabilizer code $\cC(\cS)$, which is initially corrupted by an error $\bE^{(1)}\in\{I,X,Y,Z\}^n$. Subsequently, a set of $m$ stabilizers $\{\bg_i\}_{i=1}^m$ are repeatedly measured in $r$ rounds. 
Let  $ {H}^{(\ell)}\in\{I,X,Y,Z\}^{m\times n}$ denote the check matrix being measured at round $\ell$.  (Note that $ {H}^{(\ell)}$ remains the same throughout; the superscript is for clarity. Our discussions can be readily extended to the scenario where $H^{(\ell)}$ varies for different~rounds.)
During the $\ell$-th round of syndrome extraction, it can be assumed that the syndrome is first extracted
by flawless quantum gates but the syndrome bits are subsequently subject to bit-flip errors $\be^{(\ell)}\in\{0,1\}^m$. The observed syndrome outcomes are denoted as $\bs^{(\ell)}\in\{0,1\}^m$ for $\ell=1,\dots, r$. Additionally, after the $\ell$-th round of syndrome extraction, the codeword experiences an error $\bE^{(\ell+1)}\in\{I,X,Y,Z\}^n$ for $\ell=1,\dots,r-1$.

\subsection{Phenomenological  decoding problem} 

In this error model, data qubit errors  accumulate 
and the syndrome extraction process  reflects this accumulation. 
Let $\bF^{(\ell)}\in\{I,X,Y,Z\}^n$ represent the accumulated data errors before the $\ell$-th round of syndrome extraction with $\bF^{(1)} = \bE^{(1)}$. Consequently, we have  the following equations:
\begin{samepage}
    \begin{align} 
	\bF^{(\ell)} &= \prod_{j=1}^{\ell} \bE^{(j)}  ~\in\{I,X,Y,Z\}^n,  \label{eq:F_iter} \\
	\bs^{(\ell)} &= \bF^{(\ell)}* {H}^{(\ell)}  + \be^{(\ell)} ~ \in\{0,1\}^m \label{eq:F_synd} 
    \end{align}
for $\ell=1,\dots, r$.
\end{samepage}

\begin{definition}{\bf (Phenomenological decoding problem)} \label{def:pdp}
Given a check matrix $ {H}\in\{I,X,Y,Z\}^{m\times n}$
and  $r$ rounds of syndrome extraction
 outcomes $\bs^{(\ell)}\in\{0,1\}^m$ 
 of errors $\bE^{(\ell)}\in\{I,X,Y,Z\}^{n}$ and $\be^{(\ell)}\in \{0,1\}^m$ for $\ell=1,\dots, r$, output  $\hat{\bE}^{(\ell)}\in \{I,X,Y,Z\}^{n}$ and  $\hat{\be}^{(\ell)}\in \{0,1\}^m$ for $\ell=1,\dots, r$
such that $\bs^{(\ell)} = \left( \prod_{j=1}^{\ell} \bE^{(j)}  \right)* {H}^{(\ell)} + \be^{(\ell)} ~ \in\{0,1\}^m$ 
and $ \prod_{\ell=1}^r  \bE^{(\ell)}\hat{\bE}^{(\ell)}\in  \hm( \cS)$. 
\qed
\end{definition}

To solve this decoding problem using BP, one must establish a corresponding Tanner graph. It is evident that during each round of syndrome extraction, ~(\ref{eq:F_synd}) takes on the same form as ~(\ref{eq:nos_s_i}). Consequently, we can define a local Tanner graph based on the DS matrix of $ {H}^{(\ell)}$, where variable nodes correspond to $\bF^{(\ell)}=(F^{(\ell)}_1, \dots, F^{(\ell)}_n)$ and check nodes correspond to $\bs^{(\ell)}=(s^{(\ell)}_1, \dots, s^{(\ell)}_m)$. 

The challenge remains in connecting these local Tanner graphs in a logical manner.
To address this, we introduce variable nodes corresponding to $\bE^{(\ell)}= (E^{(\ell)}_1, \dots, E^{(\ell)}_n)$  and check nodes that represent the relationships expressed by:
\begin{align} 
	\bF^{(\ell)} &= \bF^{(\ell-1)} \bE^{(\ell)}  ~\in\{I,X,Y,Z\}^n. \label{eq:F_iter2}
\end{align} 
 As a result, we create a   Tanner graph  for this decoding problem.
Figure~\ref{fig:DS_rep_F} depicts the Tanner graph for the check matrices 
	$ {H}^{(\ell)} =  
	\left[\begin{smallmatrix}
	X & Y & I \\ Z & Z & Y
	\end{smallmatrix}\right]$ 
 with three rounds of syndrome extraction.

This Tanner graph presents two challenges that make BP decoding complex:
Firstly, there is a high number of dangling  variable nodes of degree one corresponding to $e^{(\ell)}_i$ and~$E^{(\ell)}_j$.
Secondly, we lack a satisfactory strategy to initialize the variables $\bF^{(\ell)}$ for $\ell\geq 2$.
To address these issues, we derive a  decoding problem, called the generalized data-syndrome (GDS) decoding problem,  equivalent to the phenomenological decoding problem defined in Definition~\ref{def:pdp}  in the sense that solutions to one problem are also solutions to the other.

\begin{definition} \label{def:gdsp}
\noindent{\bf (GDS decoding problem)}: 
Given a check matrix $ {H}\in\{I,X,Y,Z\}^{m\times n}$
and  $r$ rounds of syndrome extraction
 outcomes $\bs^{(\ell)}\in\{0,1\}^m$ 
 of errors $\bE^{(\ell)}\in\{I,X,Y,Z\}^{n}$ and $\be^{(\ell)}\in \{0,1\}^m$ for $\ell=1,\dots, r$, output  $\hat{\bE}^{(\ell)}\in \{I,X,Y,Z\}^{n}$ and  $\hat{\be}^{(\ell)}\in \{0,1\}^m$ for $\ell=1,\dots, r$
such that 
 \begin{align}
      (\hat \bE,\hat \be)* \tilde{ {H}}'   = \bs' \label{eq:inn_H'}
 \end{align} 
and $\prod_{\ell=1}^r  \bE^{(\ell)}\hat{\bE}^{(\ell)}\in \hm(\cS)$,
where 
\begin{align}
    (\hat{\bE},\hat{\be}) =& (\hat{\bE}^{(1)},\dots, \hat{\bE}^{(r)},\, 
    \hat{\be}^{(1)},\dots, \hat{\be}^{(r)} ),\\
    \bs' =& (\bs^{(1)},\, \bs^{(1)}+\bs^{(2)},\,   \dots,\, \bs^{(r-1)} +\bs^{(r)}),\\
    \tilde{ {H}}' =& 
    \left[\begin{matrix}
    H   &\bNull&      &\bNull\\ 
    \bNull&H   &      &\bNull\\
          &      &\ddots&      \\
    \bNull&\bNull&      &H   \\
    \end{matrix}~\middle|~
    \begin{matrix}
    I_m &\bNull&      &\bNull\\ 
    I_m & I_m &      &\bNull\\
          &\ddots&\ddots&      \\
    \bNull&\bNull&I_m &I_m \\
    \end{matrix}\right]. \label{eq:gen_DS}
\end{align}

\qed
\end{definition} 
Note that the empty entries in a matrix should be interpreted as identity Pauli matrices if they correspond to $\{I,X,Y,Z\}$ 
and as zeros if they correspond to $\{0,1\}$.

The matrix $\tilde{H}'$ is called a  \textit{GDS check matrix}
and we say that $\tilde{H}'$ defines a \textit{generalized quantum DS code}.  Equation (\ref{eq:gen_DS}) can be compared to~(\ref{eq:DS_check}).
If there is only a single round of syndrome extraction, a GDS decoding problem  reduces to a DS decoding problem.

\begin{lemma} \label{lm:pdp_gdsp}
        The two decoding problems defined in Definitions~\ref{def:pdp} and~\ref{def:gdsp} are equivalent.
\end{lemma}

\begin{figure}
	\centering\resizebox{0.48\textwidth}{!}{\begin{tikzpicture}[node distance=1.3cm,>=stealth,bend angle=45,auto]

\tikzstyle{chk}=[rectangle,thick,draw=black!75,fill=black!20,minimum size=4mm]
\tikzstyle{var}=[circle,thick,draw=blue!75,fill=gray!20,minimum size=4mm,font=\footnotesize]
\tikzstyle{VAR}=[circle,thick,draw=blue!75,fill=blue!20,minimum size=5mm,font=\footnotesize]
\tikzstyle{fac}=[anchor=west,font=\footnotesize]
\tikzstyle{lin}=[rectangle,thick,draw=black!75,fill=black!75,minimum size=3mm]

\node[var] (F1^1) at (0.0,2.4) {$F^{(1)}_1$};
\node[var] (F2^1) at (1.5,2.4) {$F^{(1)}_2$};
\node[var] (F3^1) at (3.0,2.4) {$F^{(1)}_3$};
\node[chk] (z1^1) at (0.0,1.1) {};
\node[chk] (z2^1) at (1.5,1.1) {};
\node[var] (e1^1) at (0.0,0.0) {$e^{(1)}_1$};
\node[var] (e2^1) at (1.5,0.0) {$e^{(1)}_2$};

\draw[thick]                (F1^1) -- (z1^1);
\draw[thick,dashed]         (F2^1) -- (z1^1);
\draw[thick,densely dotted] (F1^1) -- (z2^1);
\draw[thick,densely dotted] (F2^1) -- (z2^1);
\draw[thick,dashed]         (F3^1) -- (z2^1);
\draw[thin]                 (z1^1) -- (e1^1);
\draw[thin]                 (z2^1) -- (e2^1);

%
%

\node[var] (F1^2) at (0.0+5,2.4) {$F^{(2)}_1$};
\node[var] (F2^2) at (1.5+5,2.4) {$F^{(2)}_2$};
\node[var] (F3^2) at (3.0+5,2.4) {$F^{(2)}_3$};
\node[chk] (z1^2) at (0.0+5,1.1) {};
\node[chk] (z2^2) at (1.5+5,1.1) {};
\node[var] (e1^2) at (0.0+5,0.0) {$e^{(2)}_1$};
\node[var] (e2^2) at (1.5+5,0.0) {$e^{(2)}_2$};

\draw[thick]                (F1^2) -- (z1^2);
\draw[thick,dashed]         (F2^2) -- (z1^2);
\draw[thick,densely dotted] (F1^2) -- (z2^2);
\draw[thick,densely dotted] (F2^2) -- (z2^2);
\draw[thick,dashed]         (F3^2) -- (z2^2);
\draw[thin]                 (z1^2) -- (e1^2);
\draw[thin]                 (z2^2) -- (e2^2);

\node[lin] (a1^2) at (0.0+2.5,2.4+1.2) {};
\node[lin] (a2^2) at (1.5+2.5,2.4+1.2) {};
\node[lin] (a3^2) at (3.0+2.5,2.4+1.2) {};
\node[var] (E1^2) at (0.0+2.5,2.4+2.4) {$E^{(2)}_1$};
\node[var] (E2^2) at (1.5+2.5,2.4+2.4) {$E^{(2)}_2$};
\node[var] (E3^2) at (3.0+2.5,2.4+2.4) {$E^{(2)}_3$};

\draw[thin] (F1^1) -- (a1^2);	\draw[thin] (E1^2) -- (a1^2);	\draw[thin] (F1^2) -- (a1^2);
\draw[thin] (F2^1) -- (a2^2);	\draw[thin] (E2^2) -- (a2^2);   \draw[thin] (F2^2) -- (a2^2);
\draw[thin] (F3^1) -- (a3^2);	\draw[thin] (E3^2) -- (a3^2);   \draw[thin] (F3^2) -- (a3^2);

\node[var] (F1^3) at (0.0+10,2.4) {$F^{(3)}_1$};
\node[var] (F2^3) at (1.5+10,2.4) {$F^{(3)}_2$};
\node[var] (F3^3) at (3.0+10,2.4) {$F^{(3)}_3$};
\node[chk] (z1^3) at (0.0+10,1.1) {};
\node[chk] (z2^3) at (1.5+10,1.1) {};
\node[var] (e1^3) at (0.0+10,0.0) {$e^{(3)}_1$};
\node[var] (e2^3) at (1.5+10,0.0) {$e^{(3)}_2$};

\draw[thick]                (F1^3) -- (z1^3);
\draw[thick,dashed]         (F2^3) -- (z1^3);
\draw[thick,densely dotted] (F1^3) -- (z2^3);
\draw[thick,densely dotted] (F2^3) -- (z2^3);
\draw[thick,dashed]         (F3^3) -- (z2^3);
\draw[thin]                 (z1^3) -- (e1^3);
\draw[thin]                 (z2^3) -- (e2^3);

\node[lin] (a1^3) at (0.0+7.5,2.4+1.2) {};
\node[lin] (a2^3) at (1.5+7.5,2.4+1.2) {};
\node[lin] (a3^3) at (3.0+7.5,2.4+1.2) {};
\node[var] (E1^3) at (0.0+7.5,2.4+2.4) {$E^{(3)}_1$};
\node[var] (E2^3) at (1.5+7.5,2.4+2.4) {$E^{(3)}_2$};
\node[var] (E3^3) at (3.0+7.5,2.4+2.4) {$E^{(3)}_3$};

\draw[thin] (F1^2) -- (a1^3);	\draw[thin] (E1^3) -- (a1^3);	\draw[thin] (F1^3) -- (a1^3);
\draw[thin] (F2^2) -- (a2^3);	\draw[thin] (E2^3) -- (a2^3);   \draw[thin] (F2^3) -- (a2^3);
\draw[thin] (F3^2) -- (a3^3);	\draw[thin] (E3^3) -- (a3^3);   \draw[thin] (F3^3) -- (a3^3);

\end{tikzpicture}}
        \caption[\CapOpn]{
		A cascade of three local Tanner graphs (c.f. Figure~\ref{fig:DS_ex1}) based on the check matrix 
			$\left[\begin{smallmatrix}
			X & Y & I \\ Z & Z & Y
			\end{smallmatrix}\right]$.
 	Notice that $\bE^{(1)} = \bF^{(1)}$ so there are no variable nodes for $\bE^{(1)}_j$.
 		A solid back box represents a check node that corresponds to the error accumulation relations described in ~(\ref{eq:F_iter2}).
	} \label{fig:DS_rep_F}
\end{figure}

\begin{proof}
Given a check matrix $ {H}\in\{I,X,Y,Z\}^{m\times n}$
and  $r$ rounds of syndrome extraction
 outcomes $\bs^{(\ell)}\in\{0,1\}^m$ 
 of errors $\bE^{(\ell)}\in\{I,X,Y,Z\}^{n}$ and $\be^{(\ell)}\in \{0,1\}^m$ for $\ell=1,\dots, r$
such that
$\bs^{(\ell)} = \left(\prod_{j=1}^\ell \bE^{(j)}\right)*  {H}^{(\ell)}  + \be^{(\ell)} ~ \in\{0,1\}^m$,
we can construct a check matrix
    \begin{align}
    \tilde{H} = 
    \left[\begin{matrix}
    H     &\bNull&      &\bNull\\ 
    H     &H     &      &\bNull\\
    \vdots&\vdots&\ddots&      \\
    H     &H     &\dots &H     \\
    \end{matrix}~\middle|~
    \begin{matrix}
    I_m &\bNull&      &\bNull\\ 
    \bNull&I_m &      &\bNull\\
          &      &\ddots&      \\
    \bNull&\bNull&      &I_m \\
    \end{matrix}\right], \label{eq:tH_meas}
    \end{align}
which satisfies the relation
\[
\bs=(\bE,\be)* \tilde{H}.
\]
This equation can be used to defines a Tanner graph without variable nodes corresponding to $\bF^{(\ell)}$. However,
the check matrix $\tilde{H}$ is dense.

Let 
    $
    R=\left[\begin{matrix}
    I_m &\bNull&      &\bNull\\ 
    I_m &I_m &      &\bNull\\
          &\ddots&\ddots&      \\
    \bNull&\bNull&I_m &I_m \\
    \end{matrix}\right]. 
    $
Then     
    \begin{align}
    \tilde{H}' &=  R\tilde{H} \notag\\ 
    &=\left[\begin{matrix}
    H     &\bNull&      &\bNull\\ 
    \bNull&H     &      &\bNull\\
          &      &\ddots&      \\
    \bNull&\bNull&      &H     \\
    \end{matrix}~\middle|~
    \begin{matrix}
    I_m &\bNull&      &\bNull\\ 
    I_m &I_m &      &\bNull\\
          &\ddots&\ddots&      \\
    \bNull&\bNull& I_m & I_m \\
    \end{matrix}\right]. \label{eq:tH_dec}
    \end{align}
is  sparser  if $H$ is sparse.
In addition, we have 
    \[
    \bs'=(\bE,\be)* \tilde{H}',
    \]
which corresponds to the GDS decoding problem in Definition~\ref{def:gdsp}.
As $R$ is of full rank, finding a solution to this GDS decoding problem also yields a solution to the phenomenological decoding problem, and vice versa.
\end{proof}

Although the two decoding problems are equivalent, the GDS  decoding problem is more suitable for BP decoding compared to the phenomenological decoding problem. 
As an illustration, consider the example of  ${H}=
	\left[\begin{smallmatrix}
	X & Y & I \\ Z & Z & Y
	\end{smallmatrix}\right]
	$
for three rounds of syndrome extraction in Figure~\ref{fig:DS_rep_F}.
The modified Tanner graph is shown in Figure~\ref{fig:DS_rep_E},
which is sparse and has no induced dangling nodes.

\begin{remark} \label{rmk:H_any}
Our method applies to the case where  different sets of stabilizer generators are measured at each round.
Suppose that  the stabilizer measurements in round $\ell$ correspond to a check matrix $H^{(\ell)}$. Since these check matrices $H^{(\ell)}$ span the same vector space, we may assume that $H^{(\ell)} = A^{(\ell)} H^{(\ell-1)}$ for some matrix $A^{(\ell)}$ for $\ell=2,\dots,r$.
Then we have a GDS check matrix
    \begin{align}
    \tilde{H} = 
    \left[\begin{matrix}
    H^{(1)} &        &       &       \\ 
    H^{(2)} &H^{(2)} &       &       \\
    \vdots  &\vdots  &\ddots &       \\
    H^{(r)} &H^{(r)} &\dots  &H^{(r)}\\
    \end{matrix}~\middle|~
    \begin{matrix}
    I &  &      &  \\ 
      &I &      &  \\
      &  &\ddots&  \\
      &  &      &I \\
    \end{matrix}\right], \label{eq:tH_meas_any}
    \end{align}
which can then be  transformed to
    \begin{align}
    \tilde{H}' = 
    \left[\begin{matrix}
    H^{(1)} &        &       &       \\ 
            &H^{(2)} &       &       \\
            &        &\ddots &       \\
            &        &       &H^{(r)}\\
    \end{matrix}~\middle|~
    \begin{matrix}
    I~~     &       &        &  \\ 
    A^{(2)} &I      &        &  \\
            &\ddots &\ddots  &  \\
            &       &A^{(r)} &I \\
    \end{matrix}\right]. \label{eq:tH_meas_dec_any}
    \end{align}
For the case of repeated syndrome measurements, it reduces to \eqref{eq:gen_DS}
with  $H^{(\ell)}=H$ and $A^{(\ell)}=I$.
If the redundant measurements can be designed flexibly,  the decoding performance can be significantly improved, even for single-shot scenarios.
Specifically, each $A^{(\ell)}$ in \eqref{eq:tH_meas_dec_any} should be chosen to be sparse. This will be discussed in Section~\ref{sec:Single_QEC}.
\qed
\end{remark}

\subsection{Lifetime of a quantum memory } \label{sec:mem_lifetime}

\begin{figure}
	\centering\resizebox{0.48\textwidth}{!}{\begin{tikzpicture}[node distance=1.3cm,>=stealth,bend angle=45,auto]

\tikzstyle{chk}=[rectangle,thick,draw=black!75,fill=black!20,minimum size=4mm]
\tikzstyle{var}=[circle,thick,draw=blue!75,fill=gray!20,minimum size=4mm,font=\footnotesize]
\tikzstyle{VAR}=[circle,thick,draw=blue!75,fill=blue!20,minimum size=5mm,font=\footnotesize]
\tikzstyle{fac}=[anchor=west,font=\footnotesize]
\tikzstyle{lin}=[rectangle,thick,draw=black!75,fill=black!75,minimum size=3mm]

\node[var] (E1^1) at (0.0,2.4) {$E^{(1)}_1$};
\node[var] (E2^1) at (1.5,2.4) {$E^{(1)}_2$};
\node[var] (E3^1) at (3.0,2.4) {$E^{(1)}_3$};
\node[chk] (z1^1) at (0.0,1.1) {};
\node[chk] (z2^1) at (1.5,1.1) {};
\node[var] (e1^1) at (0.0,0-1) {$e^{(1)}_1$};
\node[var] (e2^1) at (1.5,0-1) {$e^{(1)}_2$};

\draw[thick]                (E1^1) -- (z1^1);
\draw[thick,dashed]         (E2^1) -- (z1^1);
\draw[thick,densely dotted] (E1^1) -- (z2^1);
\draw[thick,densely dotted] (E2^1) -- (z2^1);
\draw[thick,dashed]         (E3^1) -- (z2^1);
\draw[thin]                 (z1^1) -- (e1^1);
\draw[thin]                 (z2^1) -- (e2^1);

\node[var] (E1^2) at (0.0+5,2.4) {$E^{(2)}_1$};
\node[var] (E2^2) at (1.5+5,2.4) {$E^{(2)}_2$};
\node[var] (E3^2) at (3.0+5,2.4) {$E^{(2)}_3$};
\node[chk] (z1^2) at (0.0+5,1.1) {};
\node[chk] (z2^2) at (1.5+5,1.1) {};
\node[var] (e1^2) at (0.0+5,0-1) {$e^{(2)}_1$};
\node[var] (e2^2) at (1.5+5,0-1) {$e^{(2)}_2$};

\draw[thick]                (E1^2) -- (z1^2);
\draw[thick,dashed]         (E2^2) -- (z1^2);
\draw[thick,densely dotted] (E1^2) -- (z2^2);
\draw[thick,densely dotted] (E2^2) -- (z2^2);
\draw[thick,dashed]         (E3^2) -- (z2^2);
\draw[thin]                 (z1^2) -- (e1^2);
\draw[thin]                 (z2^2) -- (e2^2);

\draw[thin]                 (z1^2) -- (e1^1);
\draw[thin]                 (z2^2) -- (e2^1);

\node[var] (E1^3) at (0.0+10,2.4) {$E^{(3)}_1$};
\node[var] (E2^3) at (1.5+10,2.4) {$E^{(3)}_2$};
\node[var] (E3^3) at (3.0+10,2.4) {$E^{(3)}_3$};
\node[chk] (z1^3) at (0.0+10,1.1) {};
\node[chk] (z2^3) at (1.5+10,1.1) {};
\node[var] (e1^3) at (0.0+10,0-1) {$e^{(3)}_1$};
\node[var] (e2^3) at (1.5+10,0-1) {$e^{(3)}_2$};

\draw[thick]                (E1^3) -- (z1^3);
\draw[thick,dashed]         (E2^3) -- (z1^3);
\draw[thick,densely dotted] (E1^3) -- (z2^3);
\draw[thick,densely dotted] (E2^3) -- (z2^3);
\draw[thick,dashed]         (E3^3) -- (z2^3);
\draw[thin]                 (z1^3) -- (e1^3);
\draw[thin]                 (z2^3) -- (e2^3);

\draw[thin]                 (z1^3) -- (e1^2);
\draw[thin]                 (z2^3) -- (e2^2);

\end{tikzpicture}}
        \caption[\CapOpn]{
		The Tanner graph for the check matrix 
			$\left[\begin{smallmatrix}
			X & Y & I \\ Z & Z & Y
			\end{smallmatrix}\right]$
   with three rounds of syndrome extraction.
		This Tanner graph depicts a decoding problem that is equivalent to the one depicted  in Figure~\ref{fig:DS_rep_F}.
	} \label{fig:DS_rep_E}
\end{figure}

Next let us consider the lifetime of a quantum memory protected by a quantum stabilizer code against phenomenological errors. 
Suppose that a round of syndrome extraction consumes a unit of time.
This assumption holds facilitate parallel syndrome measurements, allowing for syndrome measurement rounds to be completed in constant depth.

\begin{definition} \label{def:mem_time}
The \textit{lifetime} of a quantum memory protected by a stabilizer code is the average number of rounds of syndrome extraction completed before a logical error emerges on the data qubits
during the readout of the quantum memory.
\qed
\end{definition}

We assume that the readouts of a quantum memory are logical bits encoded and 
they are produced using projective measurements in the computational basis followed by a classical decoder.
This process is simpler compared to regular (non-destructive) syndrome extractions
and we presume it is error free.  This is equivalent to having a round of perfect syndrome extraction.

One can observe that  in the DS check matrix in \eqref{eq:gen_DS}, 
the last $m$  columns corresponding to the syndrome errors 
are of the form $
    \left[
    \begin{smallmatrix}
    \bZero\\ I_m
    \end{smallmatrix}\right],
    $ which shows that the last $m$ syndrome bits do not benefit from repetitive syndrome measurements. Consequently, a final round of perfect syndrome extraction   could greatly alleviate the error effects, making the decoding problem easier. Thus, the overall error correction capability can be improved.

Note that we consider the possibility of new data errors that may be introduced before the last round of perfect syndrome extraction. Figure~\ref{fig:DS_rep_E_1} extends the Tanner graph in Figure~\ref{fig:DS_rep_E}  with an additional round of perfect syndrome extraction.

\begin{figure}
	\centering\resizebox{0.5\textwidth}{!}{\begin{tikzpicture}[node distance=1.3cm,>=stealth,bend angle=45,auto]

\tikzstyle{chk}=[rectangle,thick,draw=black!75,fill=black!20,minimum size=4mm]
\tikzstyle{var}=[circle,thick,draw=blue!75,fill=gray!20,minimum size=4mm,font=\footnotesize]
\tikzstyle{VAR}=[circle,thick,draw=blue!75,fill=blue!20,minimum size=5mm,font=\footnotesize]
\tikzstyle{fac}=[anchor=west,font=\footnotesize]
\tikzstyle{lin}=[rectangle,thick,draw=black!75,fill=black!75,minimum size=3mm]

\node[var] (E1^1) at (0.0,2.4) {$E^{(1)}_1$};
\node[var] (E2^1) at (1.5,2.4) {$E^{(1)}_2$};
\node[var] (E3^1) at (3.0,2.4) {$E^{(1)}_3$};
\node[chk] (z1^1) at (0.0,1.1) {};
\node[chk] (z2^1) at (1.5,1.1) {};
\node[var] (e1^1) at (0.0,0-1) {$e^{(1)}_1$};
\node[var] (e2^1) at (1.5,0-1) {$e^{(1)}_2$};

\draw[thick]                (E1^1) -- (z1^1);
\draw[thick,dashed]         (E2^1) -- (z1^1);
\draw[thick,densely dotted] (E1^1) -- (z2^1);
\draw[thick,densely dotted] (E2^1) -- (z2^1);
\draw[thick,dashed]         (E3^1) -- (z2^1);
\draw[thin]                 (z1^1) -- (e1^1);
\draw[thin]                 (z2^1) -- (e2^1);

\node[var] (E1^2) at (0.0+5,2.4) {$E^{(2)}_1$};
\node[var] (E2^2) at (1.5+5,2.4) {$E^{(2)}_2$};
\node[var] (E3^2) at (3.0+5,2.4) {$E^{(2)}_3$};
\node[chk] (z1^2) at (0.0+5,1.1) {};
\node[chk] (z2^2) at (1.5+5,1.1) {};
\node[var] (e1^2) at (0.0+5,0-1) {$e^{(2)}_1$};
\node[var] (e2^2) at (1.5+5,0-1) {$e^{(2)}_2$};

\draw[thick]                (E1^2) -- (z1^2);
\draw[thick,dashed]         (E2^2) -- (z1^2);
\draw[thick,densely dotted] (E1^2) -- (z2^2);
\draw[thick,densely dotted] (E2^2) -- (z2^2);
\draw[thick,dashed]         (E3^2) -- (z2^2);
\draw[thin]                 (z1^2) -- (e1^2);
\draw[thin]                 (z2^2) -- (e2^2);

\draw[thin]                 (z1^2) -- (e1^1);
\draw[thin]                 (z2^2) -- (e2^1);

\node[var] (E1^3) at (0.0+10,2.4) {$E^{(3)}_1$};
\node[var] (E2^3) at (1.5+10,2.4) {$E^{(3)}_2$};
\node[var] (E3^3) at (3.0+10,2.4) {$E^{(3)}_3$};
\node[chk] (z1^3) at (0.0+10,1.1) {};
\node[chk] (z2^3) at (1.5+10,1.1) {};
\node[var] (e1^3) at (0.0+10,0-1) {$e^{(3)}_1$};
\node[var] (e2^3) at (1.5+10,0-1) {$e^{(3)}_2$};

\draw[thick]                (E1^3) -- (z1^3);
\draw[thick,dashed]         (E2^3) -- (z1^3);
\draw[thick,densely dotted] (E1^3) -- (z2^3);
\draw[thick,densely dotted] (E2^3) -- (z2^3);
\draw[thick,dashed]         (E3^3) -- (z2^3);
\draw[thin]                 (z1^3) -- (e1^3);
\draw[thin]                 (z2^3) -- (e2^3);

\draw[thin]                 (z1^3) -- (e1^2);
\draw[thin]                 (z2^3) -- (e2^2);

\node[chk] (z1^4) at (0.0+15,1.1) {};
\node[chk] (z2^4) at (1.5+15,1.1) {};
\node[var] (E1^4) at (0.0+15,2.4) {$E^{(4)}_1$};
\node[var] (E2^4) at (1.5+15,2.4) {$E^{(4)}_2$};
\node[var] (E3^4) at (3.0+15,2.4) {$E^{(4)}_3$};

\draw[thick]                (E1^4) -- (z1^4);
\draw[thick,dashed]         (E2^4) -- (z1^4);
\draw[thick,densely dotted] (E1^4) -- (z2^4);
\draw[thick,densely dotted] (E2^4) -- (z2^4);
\draw[thick,dashed]         (E3^4) -- (z2^4);

\draw[thin]                 (z1^4) -- (e1^3);
\draw[thin]                 (z2^4) -- (e2^3);

\end{tikzpicture}}
        \caption[\CapOpn]{
 The Tanner graph for the check matrix 
			$\left[\begin{smallmatrix}
			X & Y & I \\ Z & Z & Y
			\end{smallmatrix}\right]$
   with three rounds of noisy syndrome extractions and one  additional round of perfect syndrome extraction.
	} \label{fig:DS_rep_E_1} 
\end{figure}
 
 \begin{definition} \label{def:gdsp_readout}
\noindent{\bf (GDS decoding problem with readout)}: 
Given a check matrix $ {H}\in\{I,X,Y,Z\}^{m\times n}$
and  $r+1$ rounds of syndrome extraction
 outcomes $\bs^{(\ell)}\in\{0,1\}^m$ for $\ell=1,\dots r+1$
 of errors $\bE^{(\ell)}\in\{I,X,Y,Z\}^{n}$ for $\ell=1,\dots, r+1$ and $\be^{(\ell)}\in \{0,1\}^m$ for $\ell=1,\dots, r$, output  $\hat{\bE}^{(\ell)}\in \{I,X,Y,Z\}^{n}$  for $\ell=1,\dots, r+1$ and  $\hat{\be}^{(\ell)}\in \{0,1\}^m$ for $\ell=1,\dots, r$
such that 
 \begin{align}
      (\hat \bE,\hat \be)* \tilde{ {H}}''   = \bs''
 \end{align} 
and $\prod_{\ell=1}^{r+1}  \bE^{(\ell)}\hat{\bE}^{(\ell)}\in \hm(\cS)$,
where 
\begin{align}
    &(\hat{\bE},\hat{\be}) = (\hat{\bE}^{(1)},\dots, \hat{\bE}^{(r+1)},\, 
    \hat{\be}^{(1)},\dots, \hat{\be}^{(r)} ),\\
    &~~~~\, \bs'' = (\bs^{(1)},\, \bs^{(1)}+\bs^{(2)},\,   \dots,\, \bs^{(r)} +\bs^{(r+1)}),\\
    &\tilde{ {H}}'' = 
    \left[\begin{matrix}
    H   &\bNull&      &\bNull&\\ 
    \bNull&H   &      &\bNull&\\
          &      &\ddots&    &  \\
    \bNull&\bNull&      &H   &\\
       \bNull&\bNull&      &   &H\\
    \end{matrix}~\middle|~
    \begin{matrix}
    I_m &\bNull&      &\bNull\\ 
    I_m & I_m &      &\bNull\\
          &\ddots&\ddots&      \\
    \bNull&\bNull&I_m &I_m \\
     \bNull&\bNull& &I_m \\
    \end{matrix}\right]. \label{eq:gen_DS_1}
\end{align}
\qed
\end{definition}

Let $\cD_0$ and $\cD_1$ represent two QEC schemes for the GDS decoding problems in Definitions~\ref{def:gdsp} and~\ref{def:gdsp_readout}, respectively, one without readout and the other with readout, defined by a check matrix $H\in\{I,X,Y,Z\}^{m\times n}$. Performing a QEC scheme $\cD_0$ or $\cD_1$ is referred to as a \textit{QEC cycle} or simply a \textit{cycle}.

A practical quantum memory operates as follows~\cite{DKLP02}: The memory is initialized in a noiseless state. It uses QEC by $\cD_0$ based on $r$ rounds of syndrome extraction continuously and applies QEC by $\cD_1$ only when necessary (e.g., readout). 
Each QEC cycle begins with  potential residual errors from the previous cycle. 
We would like to estimate the duration for which this memory can be sustained.
For example, if, on average, the memory survives for $t$ QEC cycles and is correct after readout, its lifetime  is  $rt+1$  according to Definition~\ref{def:mem_time}, where the $+1$ corresponds to the round of readout.

Next, we will discuss how to estimate the lifetime of a quantum memory using computer simulations. In practice, QEC cycles by $\cD_0$ are continuously applied, and we do not know whether the memory will last until readout. Thus, for each QEC cycle, we conduct two parallel QEC schemes by $\cD_0$ and $\cD_1$, respectively. Here, $\cD_0$ represents the actual simulation, while $\cD_1$ checks whether the memory would survive if it were to be read out at that cycle. This process is repeated until the residual error of a QEC cycle by $\cD_1$ results in a logical error. The complete procedure is outlined in Algorithm~\ref{alg:lifetime}. 
Notice that  only the error-correction of $\cD_0$ is applied, while $\cD_1$ is a role of virtual decoder in Algorithm~\ref{alg:lifetime} to    verify whether the memory can be sustained when it is read out.
Finally, Algorithm~\ref{alg:lifetime} will be repeated multiple times, and the average of the outputs will be our estimate of the lifetime of the quantum memory.

In our simulations, the reciprocal of the lifetime will be referred to as the \textit{memory logical error rate}.

\begin{algorithm}[t]	
\begin{flushleft}
	\caption{: Sustain time} \label{alg:lifetime}
		
	{\bf Input}: a check matrix $ {H}\in\{I,X,Y,Z\}^{m\times n}$ of a stabilizer group $\cS$
and a positive integer $r$.

        {\bf Initialization}: Let:  {$\text{ROUND}=1$}. $\text{SUS}=1$. $\bE^{(0)}=I^n$.
\begin{enumerate}
    \item while ($\text{SUS}=1$)
    \begin{itemize}
		\item  Generate a set of errors
                $\bE^{(\ell)}\in\{I,X,Y,Z\}^{n}$ for $\ell=1,\dots, r+1$ 
                and $\be^{(\ell)}\in \{0,1\}^m$ for $\ell=1,\dots, r$. 
  \item Update: $\bE^{(1)}:=\bE^{(1)}\bE^{(0)}$.
  \item Generate the error syndromes  $\bs^{(\ell)}\in\{0,1\}^m$ for $\ell=1,\dots, r+1$ 
  corresponding to the errors  $\{\bE^{(1)},\dots,\bE^{(r+1)},\, \be^{(1)},\dots,\be^{(r)}\} $.
  \item Run $\cD_1$ on input $\bs^{(\ell)}\in\{0,1\}^m$ for $\ell=1,\dots, r+1$
  and output $\hat{\bE}^{(\ell)}\in\{I,X,Y,Z\}^{n}$ for $\ell=1,\dots, r+1$.

  \item 
  If $\prod_{\ell=1}^{r+1}  \bE^{(\ell)}\hat{\bE}^{(\ell)}\in \hm(\cS)$,
  \begin{itemize}
    \item Run $\cD_0$ on input $\bs^{(\ell)}\in\{0,1\}^m$ for $\ell=1,\dots, r$ 
        and output $\hat{\bE}^{(\ell)}\in\{I,X,Y,Z\}^{n}$ for $\ell=1,\dots, r$.
    \item Update $\bE^{(0)}:=\prod_{\ell=1}^{r}  \bE^{(\ell)}\hat{\bE}^{(\ell)}$.
    \item Update $\text{ROUND}=\text{ROUND}+r$;
  \end{itemize}
   Else, update $\text{SUS}=0$.

	\end{itemize}
 \item Output $\text{ROUND}$.
\end{enumerate}
 
\end{flushleft}
\end{algorithm}

\begin{algorithm}[t]	
\begin{flushleft}
  \caption{: GDS-MBP} \label{alg:DS-MBP}
	\textbf{Input}: 
		An $M'\times (N+M)$ GDS check matrix $\tilde{{ {H}}}$, where each row of $\tilde{{ {H}}}$ is in $\{I,X,Y,Z\}^{N}\times\{0,1\}^M$, 
		a~syndrome vector $\bs\in\{0,1\}^{M'}$, 
		an~integer $T_{\max}>0$, 
		a~parameter $\alpha>0$, 
		and initial LLRs ${ \{(\Lambda_j^X, \Lambda_j^Y, \Lambda_j^Z) \in \mathbb R^3\}_{j=1}^{N} }$ and
					 $\{\Lambda_{N+j}\in \mathbb R\}_{j=1}^{M}$. \\[1pt]
	{\bf Initialization.} For $j=1$ to $N+M$ and $i\in\sM(j)$:
		\begin{itemize}
		\item[] If $j\le N$, let 
			$\Gamma_{j\to i} = \lambda_{\tilde H_{ij}} (\Lambda_j^X, \Lambda_j^Y, \Lambda_j^Z)$.
		\item[] Else, let $\Gamma_{j\to i} = \Lambda_j$.
		\end{itemize}
	{\bf Horizontal Step.} For $i=1$ to $M'$ and $j\in\sN(i)$:
	\begin{align*}
	\Delta_{i\to j} = (-1)^{s_i}\underset{j'\in\sN(i)\setminus \{j\}}{\boxplus} \Gamma_{j'\to i}.   
	\end{align*}

 	{\bf Vertical Step.} For $j=1$ to $N+M$:
	\begin{itemize}
		\item[] If $j\le N$: ~
			$ \displaystyle 
			\Gamma_{j}^W = \Lambda_j^W + \frac{1}{\alpha} \sum_{i\in\sM(j): \atop W*\tilde H_{ij} =1} \Delta_{i\to j},
			$\\
			\mbox{\quad} for $W\in\{X,Y,Z\}$.

		\item[] Else: ~
		 	$ \displaystyle 
			\Gamma_{j} = \Lambda_j + \frac{1}{\alpha} \sum_{i\in\sM(j)} \Delta_{i\to j}.
			$

		\item ({\bf Hard Decision}) 
			\begin{itemize}
			\item Let $\hat{\bE} = (\hat E_1,\dots,\hat E_{N})$, 
				where\\ $\hat E_j = I$ if $\Gamma_{j}^W > 0$ for all $W\in\{X,Y,Z\}$,
    and  $\hat E_j = \argmin\limits_{W\in\{X,Y,Z\}} \Gamma_{j}^W$, otherwise.
			\item Let $\hat{\be} = (\hat e_{1},\dots,\hat e_{M})$, where\\ $\hat e_j=0$ if $\Gamma_{N+j}>0$, and  $\hat e_j=1$, otherwise.
			\end{itemize}
			\begin{itemize}
			\item If $(\hat{\bE},\hat{\be})*\tilde{{ {H}}} = \bs$, return ``CONVERGE''
   and output $(\hat{\bE},\hat{\be})$;
			\item Else, if the maximum number of iterations $T_{\max}$ is reached, halt and return ``FAIL'';
   			\item   Else, for $j=1$ to $N$ and $i\in\sM(j)$:
				\begin{itemize}[leftmargin=-1pt]
					\item[] If $j\le N$: 
			 			\begin{align*} 
			 			\Gamma_{j\to i}^W &= \Gamma_j^W - (W * \tilde H_{ij}) \Delta_{i\to j}, ~ W\in\{X,Y,Z\}, \\ 
						\Gamma_{j\to i} &= \lambda_{\tilde H_{ij}} (\Gamma_{j\to i}^X, \Gamma_{j\to i}^Y, \Gamma_{j\to i}^Z).	
			 			\end{align*}	
					\item[] Else: ~
			 			$\Gamma_{j\to i} = \Gamma_j - \Delta_{i\to j}.$
			\end{itemize}
 			\item Repeat from the horizontal step.
			\end{itemize}
		\end{itemize}

\end{flushleft}
\end{algorithm}

\section{Belief propagation decoding} \label{sec:dec}

BP has proven its effectiveness in stabilizer decoding  problems~\cite{KL21}. It has also been successfully extended to accommodate DS decoding problems in several examples~\cite{KCL21}. In this section, we take a step further by expanding BP decoding to tackle coding problems defined over mixed alphabets, including $\{I, X, Y, Z\}$ and $\{0, 1\},$ specifically addressing GDS decoding problems. 
Our algorithms, named GDS-MBP and GDS-AMBP, are based on the MBP and AMBP algorithms~\cite{KL21}.  These algorithms are presented in detail in Algorithms~\ref{alg:DS-MBP} and \ref{alg:DS-AMBP},
which will be explained as follows.

 We consider  error variables $\bE = (E_1, \dots, E_N) \in \{I, X, Y, Z\}^N$ and $\be = (e_{1}, \dots, e_{M}) \in \{0, 1\}^M$. Each Pauli variable $E_j$ is independently generated with a depolarizing rate $\epsilon \in [0, 3/4)$ following the distribution $(p_j^I, p_j^X, p_j^Y, p_j^Z) = ({1-\epsilon},\, \epsilon/3,\, \epsilon/3,\, \epsilon/3)$. Each syndrome bit error $e_j$ is an independent bit-flip with rate $\epsilon_\text{b} \in [0, 1/2)$ and follows the probability distribution $(q_j^{(0)}, q_j^{(1)}) = (1-\epsilon_\text{b}, \epsilon_\text{b}).$
Thus the error vector $(\bE,\be)\in\{I,X,Y,Z\}^N\times\{0,1\}^M$ occurs with probability
\begin{align}
	&\quad \text{Pr}{(\bE,\be)} = \prod_{j=1}^N p^{E_j} \, \prod_{j=1}^M q^{(e_{j})}  \notag \\
	&= \left(\frac{\epsilon}{3}\right)^{\wt(\bE)}(1-\epsilon)^{N-\wt(\bE)} \, \epsilon_\text{b}^{\wt(\be)}(1-\epsilon_\text{b})^{M-\wt(\be)},  \label{eq:p_E_e}
\end{align}
where $\wt(\be)$ is the Hamming weight of $\be$.

 The initial log-likelihood ratio (LLR) distribution of $E_j$ is $(\Lambda_j^X, \Lambda_j^Y, \Lambda_j^Z)$, where
	$$
	\Lambda_j^W = \ln(p_j^I/p_j^W), ~ W\in\{X,Y,Z\}, ~ j\in\{1,\dots, N\}.
	$$
Likewise, the initial LLR distribution of $e_j$ is 
	$$
	\Lambda_{N+j} = \ln(q_j^{(0)}/q_j^{(1)}), ~ j\in\{1,\dots,M\}.
	$$

 Suppose that we have a GDS check matrix $\tilde{H}$ of size $M'\times (N+M),$ where the first $N$ columns use the Pauli operators $\{I,X,Y,Z\}$, and the remaining $M$ columns use a binary alphabet $\{0,1\}$. 
 This $\tilde{H}$ naturally induces a Tanner graph, as explained in the previous section.

 Given a syndrome vector $\bs\in\{0,1\}^{M'}$ for $(\bE,\be)$, where  
 \begin{align}
     \bs= (\bE,\be)*\tilde{H},\label{eq:check_relation}
 \end{align}
BP operates through iterative message-passing on the Tanner graph to continuously have the updated LLR distribution $\Gamma_j^X, \Gamma_j^Y, \Gamma_j^Z$ (for $1\le j\le N$) and $\Gamma_{N+j}$ (for $ 1\leq j\leq M$) for each $E_j$ and $e_j$.
It can be demonstrated that the calculations in GDS-MBP provide an approximation of the conditional marginal distribution for each error node~\cite{KL21a}.
The details are provided in the Appendix.

At each iteration, we compute variable-to-check messages $\Gamma_{j\to i}$ and check-to-variable messages $\Delta_{i\to j}$. We make a hard decision on the error estimate at each iteration, and if the syndrome is matched, the process halts. If the syndrome is not matched, the iterative process may continue for a maximum of $T_{\max}$ iterations before declaring a failure.

Let the set of neighboring nodes for a check node $i$ be denoted as 
    $\sN(i) = \{j: \tilde H_{ij}\in\{X,Y,Z\} \text{ or } \tilde H_{ij}=1\}.$ 
Similarly, for a variable node $j$, we have 
    $\sM(j) = \{i: \tilde H_{ij}\in\{X,Y,Z\} \text{ or } \tilde H_{ij}=1\}$.
Given ~(\ref{eq:check_relation}) and the LLR distributions, which collectively indicate
whether   $E_j$ is more likely to commute or anticommute with an entry of $\tilde{H}$, 
we only require a single LLR value. 
 To achieve this,  we use a function $\lambda_W:\mR^3\rightarrow \mR$ for  each $W\in\{X,Y,Z\}$ defined as
	\begin{align} 
	\lambda_W(\gamma^X,\gamma^Y,\gamma^Z) = \ln \frac{1+ e^{-\gamma^{W}}}{ e^{-\gamma^{X}}+e^{-\gamma^{Y}}+e^{-\gamma^{Z}}-e^{-\gamma^{W}} }. \label{eq:la}
	\end{align} 
As a result, only scalar messages are transmitted on the Tanner graph induced by $\tilde{H}$. 

The check-node computation is performed using  the operator $\boxplus$ defined by: for a set of $\ell$ real scalars $a_1,a_2,\dots,a_\ell \in \mR$,
	\begin{equation} \label{eq:bsum} 
	\overset{\ell}{\underset{j=1}{\boxplus}} \, a_j = 2\tanh^{-1} \textstyle \left( \prod_{j=1}^\ell \displaystyle \tanh\frac{a_j}{2} \right).
	\end{equation}

We will discuss $\lambda_W$ and $\boxplus$ in Appendix.

 GDS-MBP, inheriting from MBP, includes a parameter $\alpha$ that controls the step size in message passing. This parameter   can be determined based on the average weight of the rows of the check matrix and the physical error rates, with preliminary simulations (see Appendix B.1 of the arXiv version of \cite{KL21}). 
 For improved performance, an optimal $\alpha^*$ can be  adaptively selected.
 GDS-AMBP,  provided in Algorithm~\ref{alg:DS-AMBP}, is such an adaptive scheme that extends GDS-MBP.

\begin{algorithm}[t]	
\begin{flushleft}
	\caption{: GDS-AMBP} \label{alg:DS-AMBP}
		
	{\bf Input}: An $M'\times (N+M)$ GDS check matrix $\tilde{{ {H}}}$, where each row of $\tilde{{ {H}}}$ is in $\{I,X,Y,Z\}^{N}\times\{0,1\}^M$, 
		a~syndrome vector $\bs\in\{0,1\}^{M'}$, 
		an~integer $T_{\max}>0$, 
  	a~sequence of $\ell$ parameters $\alpha_1 > \alpha_2 >\dots > \alpha_\ell > 0$,
		and initial LLRs ${ \{(\Lambda_j^X, \Lambda_j^Y, \Lambda_j^Z) \in \mathbb R^3\}_{j=1}^{N} }$ and
					 $\{\Lambda_{N+j}\in \mathbb R\}_{j=1}^{M}$. \\[1pt]
 
	{\bf Initialization}:  Let $i=1$. 

\begin{itemize}
     \item {\bf BP Step}: Run GDS-MBP$(\tilde{{ {H}}},\, \bs,\, T_{\max},\, \alpha_i,\, \text{initial LLRs})$,
	\begin{itemize}
		\item[] which returns  ``CONVERGE'' or ``FAIL'' with $(\hat \bE, \hat \be) \in \{I,X,Y,Z\}^{N}\times\{0,1\}^M$.
	\end{itemize}
    \item  {\bf Adaptive Check}: 
	\begin{itemize}
		\item If ``CONVERGE'', output  $(\hat \bE, \hat \be)$ and $\alpha^* = \alpha_i$ and return ``SUCCESS'';
		\item Else, if $i<\ell$, update $i:= i+1$ and repeat from the BP Step;
		\item Else, return ``FAIL''.
	\end{itemize}
\end{itemize}

\end{flushleft}
\end{algorithm}

In the following subsections, we will discuss some techniques for GDS matrix design.

\section{Phenomenological quantum memory} \label{sec:mem_QEC}

In this section, we analyze 2D topological quantum codes used in the context of phenomenological quantum memory. These codes are designed with stabilizers that are intentionally local and have low weight. Importantly, a full set of stabilizers can be measured in a single, simultaneous round of syndrome extraction. As a result, these codes are particularly well-suited to the phenomenological noise model.

Instead of analyzing  only the block-wise logical error rate, we will conduct a comprehensive simulation of quantum  memory lifetime by~Algorithm\,\ref{alg:lifetime}. 
In our simulations, the decoders $\cD_0$ and $\cD_1$ can be implemented using either   GDS-MBP (Algorithm~\ref{alg:DS-MBP})  or GDS-AMBP (Algorithm~\ref{alg:DS-AMBP}).

\subsection{Complexity of GDS-(A)MBP for topological codes} \label{sec:F_err}

\begin{figure}
    \centering
    \subfloat[\label{fig:toric_2x2}]{\includegraphics[width=0.12\textwidth]{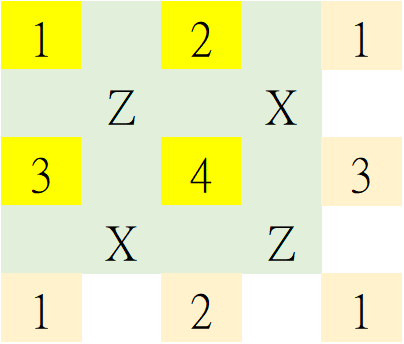}} ~~~~~~~~ 
    \subfloat[\label{fig:toric_4x4}]{\includegraphics[width=0.16\textwidth]{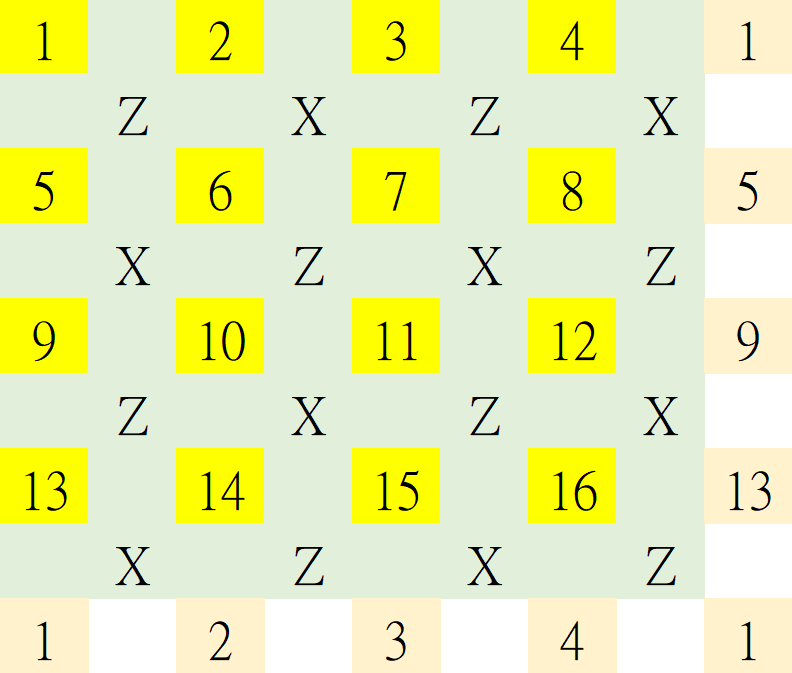}}
    \caption{
        The lattice representation of the family of $[[L^2, 2, L]]$ toric codes: (a)~${L=2}$ and (b)~$L=4$.
        Each data qubit is represented by a yellow box labeled with a number ranging from 1 to $L^2$. The orange boxes on the boundary correspond to the yellow boxes with the same numbers,  indicating the connectivity of qubits due to the toroidal topology of the lattice.
The green boxes, labeled with either $X$ or $Z$ and surrounded by four data qubits (say $i$, $j$, $k$, and $l$), represent stabilizers of the form $X_i X_j X_k X_l$ or $Z_i Z_j Z_k Z_l$, respectively.
    } \label{fig:toric_lat}		
    \end{figure}

In the phenomenological noise model, achieving effective error correction with topological codes necessitates multiple rounds of syndrome extraction. In our simulations, we will focus on the family of $[[n=L^2,\, k=2,\, d=L]]$ rotated toric codes for even integers $L$ (as defined in \cite{BM07,HFDM12}). 
The lattices of the code for $L=2$ and $L=4$ are depicted in Figure~\ref{fig:toric_lat}.
Our experiments indicate that to achieve good decoding performance, up to $2L$ rounds of syndrome extraction are required. Furthermore, the decoding performance starts to saturate even with more rounds of syndrome extraction, up to $4L$.
We conjecture that this holds for general  2D topological quantum codes.

For a $n$-qubit topological code with a minimum distance of $d$, each round of syndrome measurements involves $m=O(n)$ stabilizer measurements. We can reasonably assume that approximately $O(d)$ rounds of repeated measurements are needed. Consequently, the GDS check matrix will contain $N=O(nd)$ quaternary error variables and $M=O(md) =O(nd)$ binary error variables. Since the minimum distance of an $n$-qubit 2D topological code scales as $O(n^{0.5})$, the GDS check matrix will thus have $O(N+M)=O(n^{1.5})$ error variables.

Similar to conventional BP, GDS-MBP has a worst-case time complexity of $O((N+M)\gamma T_{\max})$ (cf.~\cite{KL21a,KL21}), where $\gamma$ is the mean column-weight of the GDS check matrix, and $T_{\max}$ is the maximum number of iterations.
For BP performance, $T_{\max}$ is  chosen to be $O(\log N)$,
which is generally sufficient. 
In the case of  a  topological code, the mean column-weight of its GDS check matrix is roughly a constant, independent of the code length. 
Therefore, the complexity of GDS-MBP is $O((N+M) \log N) = O(n^{1.5}\log n)$.

On the other hand, GDS-AMBP runs with a linear sweep  from $\alpha_1$ to $\alpha_0$ with decrement $0.01$, where $\alpha_1>\alpha_0>0$ and $\alpha_1$ is linearly proportional to the mean row-weight $\rho$ of the GDS check matrix~\cite[Figure~3 of the arXiv version]{KL21}. 
Based on our experience, a decrement of 0.01 is sufficient, as 
$\alpha$ controls the strength of message passing, and BP usually converges in 
 $O(\log N)$ iterations, making it relatively insensitive.
Therefore, the complexity of GDS-AMBP is
$O( (N+M) \rho \gamma T_{\max})$. 
For 2D topological codes,  $\rho$ is also  roughly a  constant, independent of the code length. Thus the complexity reduces to  $O((N+M)\log N) = O(n^{1.5}\log n)$.

We can compare the complexities of other decoders for reference.  The RG decoder exhibits a complexity of $O(N\log N)= O(n^{1.5}\log n)$ as well. The MWPM has a complexity of $O(N^3)=O(n^{4.5})$, or $O(N^2)=O(n^3)$ when local matching is employed. These numbers are summarized in Table~\ref{tb:toric_ph}.

\subsection{Simulations on toric codes} \label{sec:sim_toric}

In this section, we conduct simulations on a quantum memory implemented with a code from a family of $[[L^2,2,L]]$ rotated toric codes for even integers $L$. In this phenomenological noise model, we assume that the syndrome error rate and the data error rate are equal.\footnote{
    Assuming every location in a circuit has a physical error rate $p$, with errors not propagating through CNOT gates, the error distribution for a two-qubit CNOT gate is $\text{Pr}(E_1,E_2)=p/15$ for $(E_1,E_2)\in\{I,X,Y,Z\}^2\setminus \{(I,I)\}$ and $\text{Pr}(I,I)=1-p$. This leads to the marginal distribution: $\text{Pr}(E_1\ne I) = 12p/15$ and $\text{Pr}(E_1=X) = 8p/15$. In a round of syndrome extraction for the toric code, each data qubit is involved in four stabilizer measurements using four CNOT gates. Thus, the data error rate is $\epsilon = (1 + 4\times12/15)p = 4.2p$.

    Each syndrome bit extraction uses an ancillary qubit, four CNOT gates, and one single-qubit measurement. Therefore, the syndrome error rate is $\epsilon_\text{b} = (2 + 4\times 8/15)p = 4.1333p$, close to the data error rate. In the phenomenological noise model, we may simply assume $\epsilon=\epsilon_\text{b}$.

    If this model is addressed by decoding $X$ and $Z$ errors separately (e.g., using MWPM and RG), then the bit-flip error rate $\epsilon_X$ is $2\epsilon/3$ (see \cite{MMM04}). Since $\epsilon=\epsilon_\text{b}$, we have $\epsilon_\text{b}=1.5\epsilon_X$. However, for simplicity, it is commonly assumed that $\epsilon_\text{b}=\epsilon_X$ in the  simulations of MWPM and RG \cite{WHP03,Har04, DP14}.

    }

To estimate the memory lifetime (Definition~\ref{def:mem_time}), we run Algorithm~\ref{alg:lifetime}, wherein each decoder $\cD_0$ or $\cD_1$ can be implemented by either GDS-MBP or GDS-AMBP. 
For improved performance, we consider both $\cD_0$ and $\cD_1$ implemented by GDS-AMBP.

For BP decoding to be effective on the rotated toric codes, a non-parallel schedule and $\alpha<1$ are required~\cite{KL21}.
In the following, we will use the serial schedule (along variable nodes) for BP
and set the maximum number of iterations $T_{\max}=150$. 
This schedule can be highly parallelized as discussed in \cite[arXiv Sec.~5]{KL21}.
We simulate GDS-AMBP with $\alpha^*\in\{1.20, 1.19, \dots, 0.30\}$.

    \begin{figure*}
    \centering
    (a)~\raisebox{-.5\height}{\includegraphics[width=0.4\textwidth]{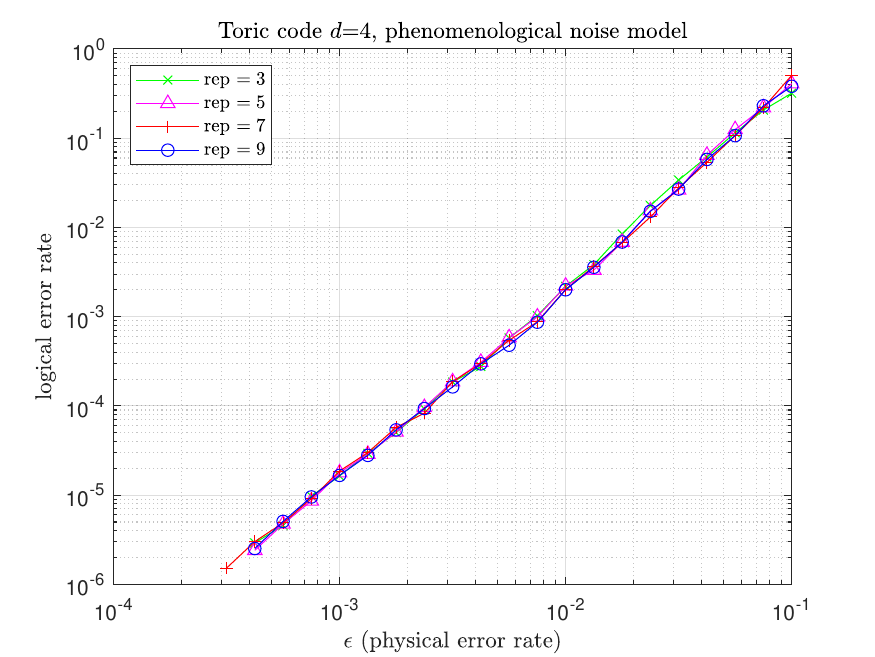}}\vspace*{\floatsep}
    (b)~\raisebox{-.5\height}{\includegraphics[width=0.4\textwidth]{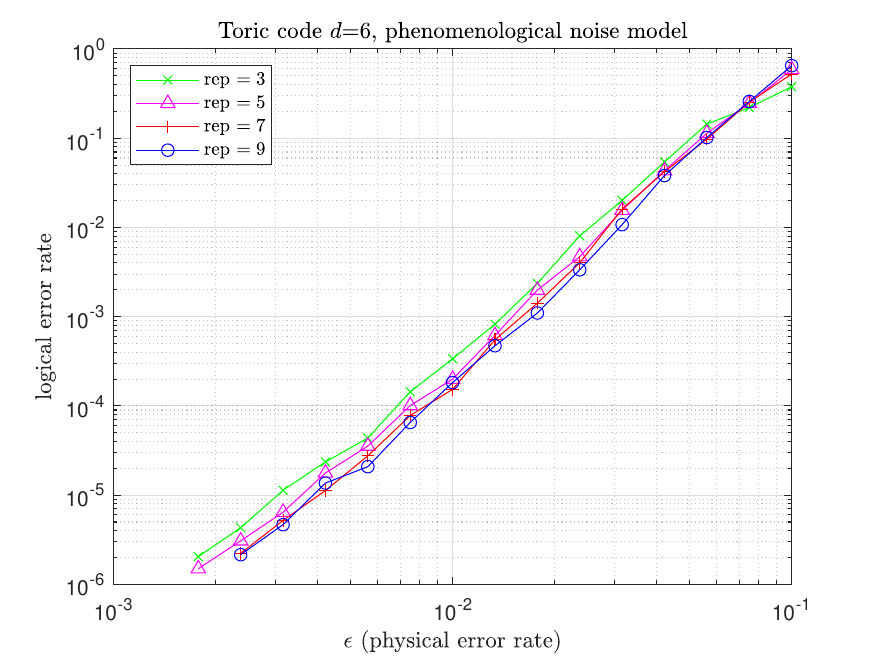}}\vspace*{\floatsep}
    (c)~\raisebox{-.5\height}{\includegraphics[width=0.4\textwidth]{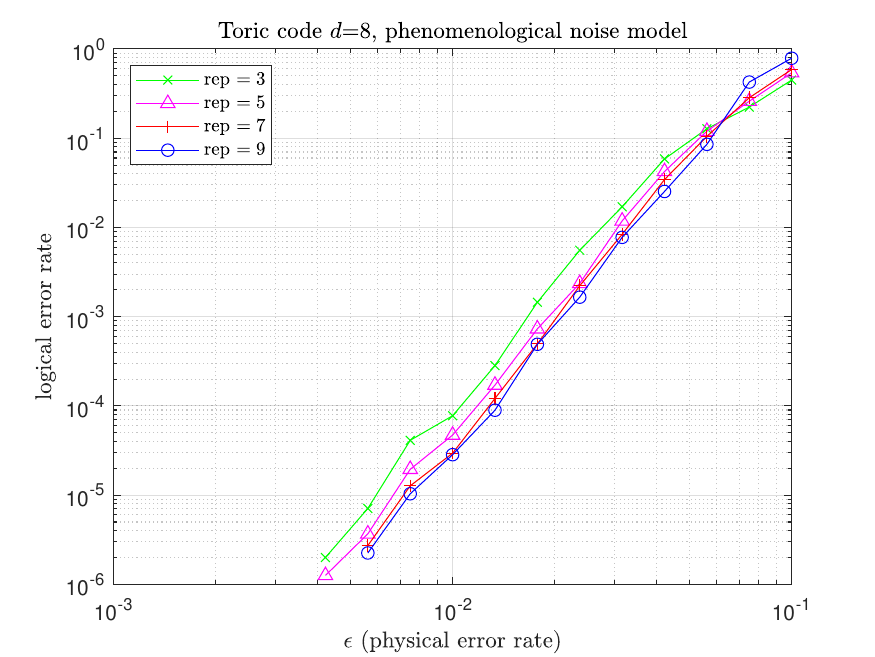}}\vspace*{\floatsep}
    (d)~\raisebox{-.5\height}{\includegraphics[width=0.4\textwidth]{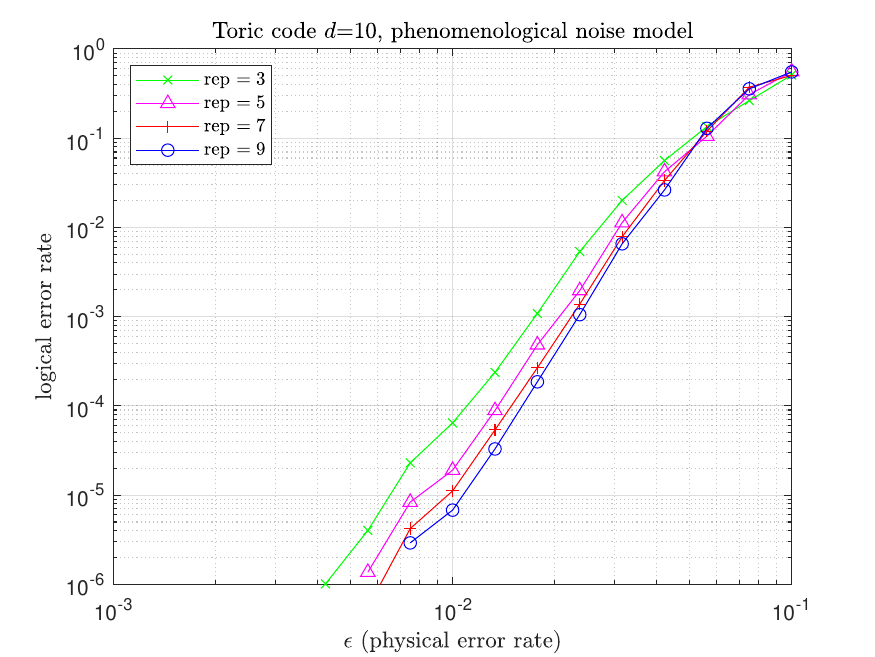}}\vspace*{\floatsep}
    (e)~\raisebox{-.5\height}{\includegraphics[width=0.4\textwidth]{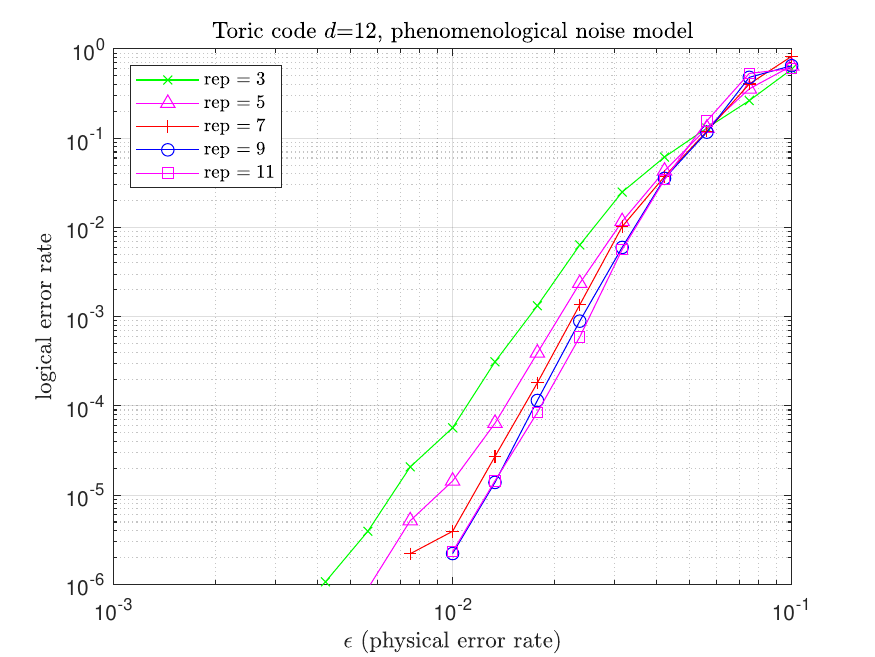}}\vspace*{\floatsep}
    (f)~\raisebox{-.5\height}{\includegraphics[width=0.4\textwidth]{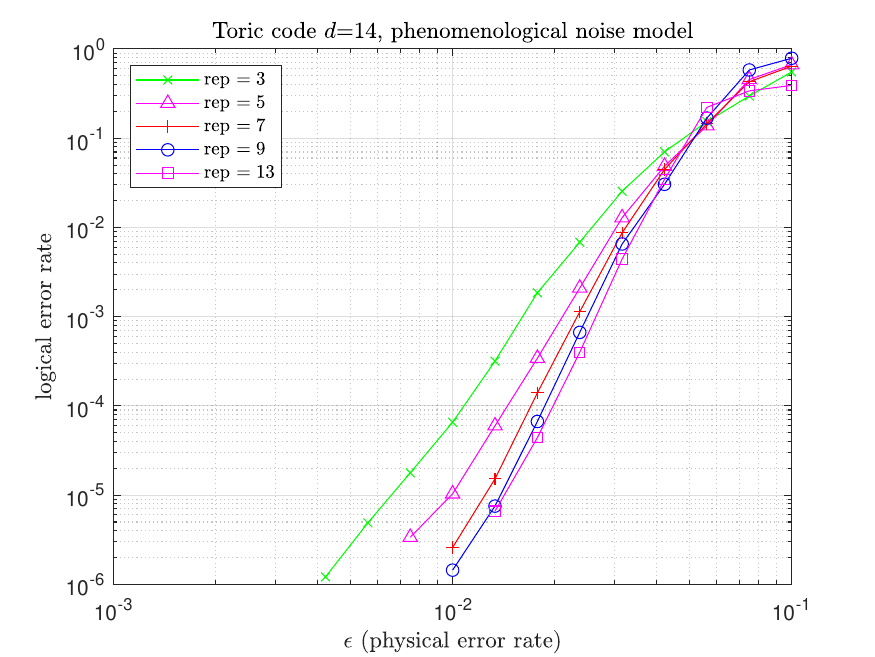}}\vspace*{\floatsep}
    (g)~\raisebox{-.5\height}{\includegraphics[width=0.4\textwidth]{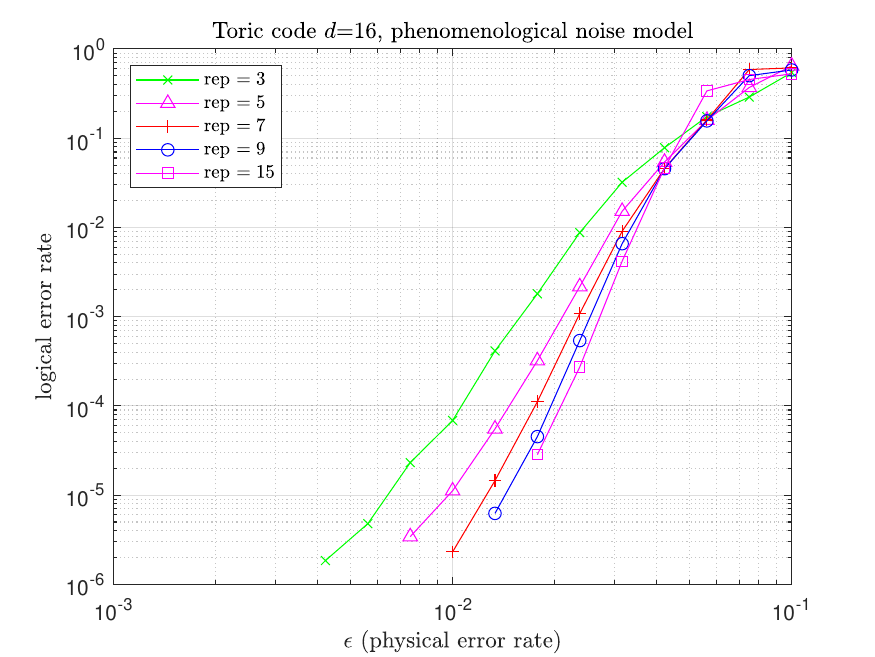}}\vspace*{\floatsep}
    (h)~\raisebox{-.5\height}{\includegraphics[width=0.4\textwidth]{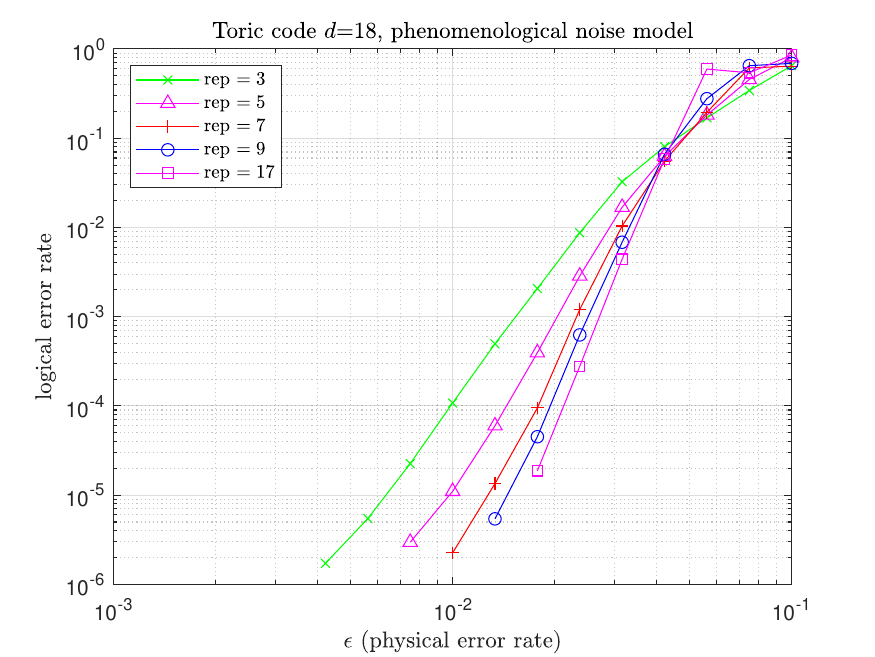}}
    \caption{
		The performances of GDS-AMBP in correcting quantum memories protected by the rotated toric codes: (a)~$d=4$, (b)~$d=6$, (c)~$d=8$, (d)~$d=10$, (e)~$d=12$, (f)~$d=14$, (g)~$d=16$, and (h)~$d=18$.
    } \label{fig:toric_all}		
    \end{figure*}

In Figure~\ref{fig:toric_all}, we show the performance of GDS-AMBP in correcting quantum memories protected by the rotated toric codes with distances $d=6, 8, \dots, 18$, respectively. The performance of each code is simulated for various rounds ($r$) of syndrome extractions, for $r=3,5,7,9,$ and $d-1$. These are indicated by the term `rep' in the legend for clarity. We have collected at least 100 logical error events for every data point in our simulations.

  As can be seen, under high physical error rates, smaller values of $r$ tend to yield better performance. This is because a high number of syndrome-extraction rounds can accumulate errors, potentially leading to uncorrectable logical errors. Conversely, when the physical error rate is sufficiently low, a larger number of rounds results in improved performance. As previously mentioned, it is suggested to perform approximately $O(d)$ rounds of syndrome extraction. It is also evident that performance gains become less significant when $r$ exceeds~$d$. This  reflects that, around $r\approx d$, the error correction capabilities for syndrome errors and data errors begin to align.

\begin{figure}
    \centering \includegraphics[width=0.4\textwidth]{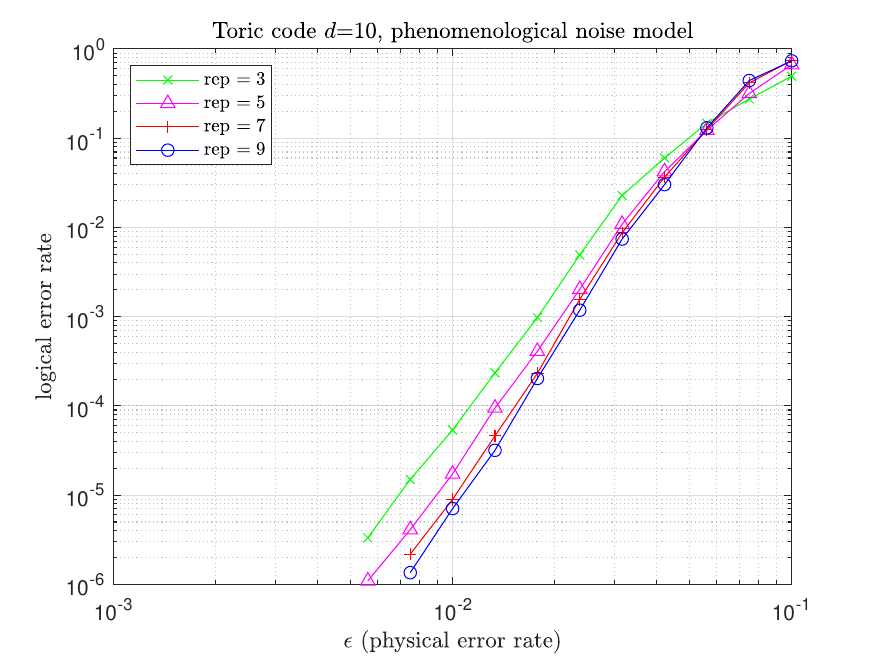}
    \caption{
    The performance of GDS-AMBP on topological toric quantum memory with $d=10$ and fixed initialization $\epsilon_0 = 0.001$. 
    } \label{fig:toric_10_fixP1}
\end{figure}

In Figure~\ref{fig:toric_all}, these curves are simulated with 
the initial probabilities of GDS-AMBP chosen according to the physical error rate $\epsilon$. It can be seen that the case $d=10$ seems to have an error floor, even when $r=d-1$. This can be improved  using  more repetitions of syndrome extractions. Alternatively, this can be improved by using fixed initialization~\cite{HFI12,KL21}. We have simulated this case using fixed initialization at $\epsilon_0=0.001$, regardless of the physical $\epsilon$, 
and the resulting performance curve is presented in Figure~\ref{fig:toric_10_fixP1}. 
As can be seen, there is no observable  error floor in the performance,
and we will use this  performance   in the following threshold analysis.

\begin{figure}
	\centering \includegraphics[width=0.5\textwidth]{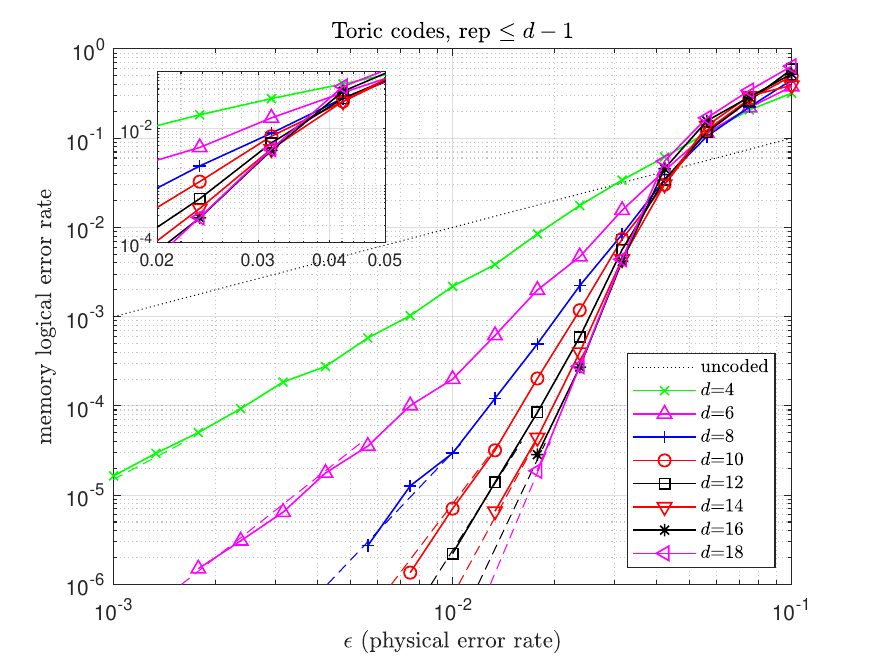}
	\caption{
 The performance of GDS-AMBP on topological toric quantum memories with varying distances $d$  in the phenomenological noise model. Each dashed curve serves as a reference for $a\epsilon^{t+1}$, where $t=\lfloor\frac{d-1}{2}\rfloor$ and $a$ is an appropriate scalar.
	} \label{fig:toric_ph}
\end{figure}

\begin{figure}
	\centering \includegraphics[width=0.5\textwidth]{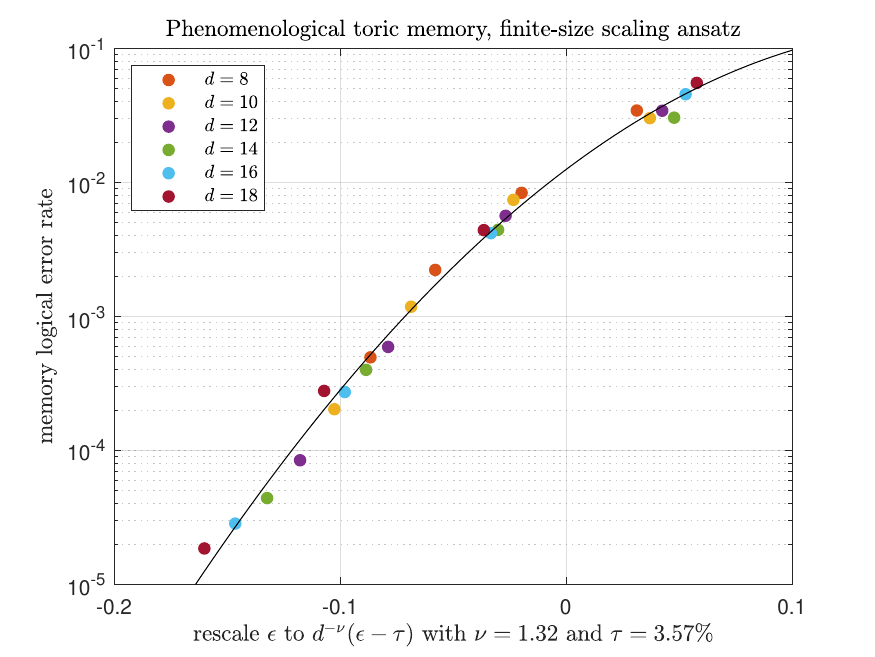}
	\caption{
 Estimation of the error threshold using the finite-size scaling ansatz. This figure represents a rescaled version of Figure~\ref{fig:toric_ph}, where $\epsilon$ is rescaled to $d^{1/\nu}(\epsilon-\tau)$.
	} \label{fig:ansatz}
\end{figure}

In Figure~\ref{fig:toric_ph}, performance curves are presented for the rotated toric codes with distances $d=4,6,\dots, 18$. Each curve for a code with distance $d$ is optimized across various rounds $r$ ranging from 3 to $d-1$. To highlight the overall performance for each $d$, one can utilize the lower envelope of the curves in each subfigure of Figure~\ref{fig:toric_all} (and Figure~\ref{fig:toric_10_fixP1} for $d=10$) as a representation of the optimized performance across different $r$ values.

Additionally, for each $d$, a dashed curve $a\epsilon^{t+1}$ is included as a reference curve, where $t=\lfloor\frac{d-1}{2}\rfloor$, and $a$ is a scalar chosen to approximate the actual performance curve. 
The reference curves illustrate that the expected error correction capabilities for each specific $d$ are achieved using GDS-AMBP.

 Next, we assess the error threshold for this family of phenomenological quantum memories using the finite-size scaling ansatz method~\cite{WHP03}. To do so, we rescale $\epsilon$ to $d^{1/\nu}(\epsilon-\tau)$ and replot  Figure~\ref{fig:toric_ph}. We seek the optimal values of $\nu\in [1, 2]$, with a precision of 0.01, and $\tau\in [0.02, 0.04]$, with a precision of 0.0001, to fit a second-order curve with minimized mean squared error. The resulting values are 
 $\nu=1.32$ and $\tau=3.57\%$. 
 The rescaled figure is presented in Figure~\ref{fig:ansatz}.

The value of $\tau$ provides an estimate of the error threshold for GDS-AMBP on toric codes within the phenomenological noise model. 
Consequently, we obtain an estimated error threshold of more than 3\%.

 \begin{table}
	\caption{ 			 
		The  thresholds  and complexities of various decoders on the toric code memory in the phenomenological noise model.
	} 					
	\label{tb:toric_ph}
	\centering 
	$\begin{array}{|l|l|l|l|}
	\hline
	\text{ noise type} &\text{decoder}							& \text{threshold} & \text{complexity}\\ 
	\hline                                              
	\text{$X$-$Z$}&\text{MWPM \cite{WHP03}} 				& 2.93\%					& O(N^3)=O(n^{4.5})             \\
	\text{independent}&\text{Recovering RPGM \cite{OAIM04}}	& 3.3\%						& --                            \\	
\text{data errors}	&\text{RG \cite{DP14}}					& 1.94\%					& O(N\log N) \\
 &&&= O(n^{1.5}\log n) \\
 	\hline
	\text{depolarizing}&\text{DS-AMBP (this paper)}				& 3.52\%			& O(N\log N)\\
 \text{data errors}&&&= O(n^{1.5}\log n) \\
	\hline
	\end{array}$
	%
	{\footnotesize \medskip
	\begin{itemize}[leftmargin=*]
        \item[] 
            Note that  $n$ represents the code length, and $N=nr=O(n^{1.5})$ refers to the number of input variables, assuming the number of repeated measurement rounds $r=O(\sqrt{n})$.
            MWPM with local matching can have a complexity of $O(N^2) = O(n^3)$, although there may be a slight performance loss to consider. The decoding complexity in \cite{OAIM04} is not specified but is based on Monte Carlo sampling, which is generally associated with high complexity.
	\end{itemize}
	}
\end{table}

For reference, Table~\ref{tb:toric_ph} in Section~\ref{sec:mem_QEC} summarizes thresholds for several other schemes in the phenomenological noise model. The optimal threshold value for independent $X$-$Z$ errors on the data qubits is estimated using a recovering random-plaquette gauge model (recovering RPGM) based on Monte Carlo sampling, which can be computationally intensive \cite{OAIM04}. Additionally, toric codes can be decoded by minimum-weight perfect matching (MWPM) \cite{Edm65} (see \cite{WHP03,Har04,WFSH10}) or renormalization group (RG) \cite{DP10}. Both MWPM and RG can be extended to handle errors in the phenomenological noise model \cite{WHP03,Har04, DP14}, but there are unsatisfying issues: the complexity of MWPM is nonlinear, while RG has a minor performance loss. Unfortunately, these decoders are only analyzed over independent $X$-$Z$ noises. Due to the nature of their algorithms, which do not consider the correlations between the $X$ and $Z$ errors, these decoders are suboptimal in independent depolarizing noise.

\section{Single-shot quantum error correction} \label{sec:Single_QEC}

In this section, we focus on a special case of the phenomenological noise model involving only a single round of noisy syndrome extraction, with redundant stabilizer measurements allowed. This scenario is referred to as single-shot quantum error correction, and it becomes particularly important when the error rate of syndrome measurements is dominant over all other errors. The single-shot performance of a QEC scheme is estimated by simulating the block-wise logical error rate using  Algorithm~\ref{alg:DS-MBP} or Algorithm~\ref{alg:DS-AMBP},
  which is simpler than quantum memory lifetime estimation that involves Algorithm~\ref{alg:lifetime}.

\subsection{Design of low-overhead DS check matrix} \label{sec:E_err}

In single-shot quantum error correction, it is common to measure redundant stabilizers to ensure that the syndromes are robust against measurement errors. By including an arbitrary number of stabilizers in the syndrome extraction, the impact of measurement errors can be minimized. However, since we are limited to a single round of syndrome extraction, it is reasonable to focus on a set of stabilizers that can be measured in parallel. Consequently, we assume only a small number of redundant stabilizers can be included in one round of syndrome extraction.

Let $H\in\{I,X,Y,Z\}^{m\times n}$ be a check matrix of a quantum code. 
Consider its DS check matrix 
    \begin{align}
    \small
    \left[\begin{array}{c|cc}
        H  &  {I}_m &          \\
        AH &        & {I}_\ell \\
    \end{array}\right],
    \label{eq:GB_meas}
    \end{align}
    where  ${A}\in\{0,1\}^{\ell\times m}$ and $AH$ represents $\ell$ additional redundant stabilizers. 
After performing  Gaussian elimination on this check matrix, we have
	\begin{equation} \label{eq:S_dec}
	\tilde{H} = \small
		\left[\begin{array}{c|cc}
		 {H}   &    {I}_m &    \\
		 &    {A}   &  {I}_\ell \\
		\end{array}\right].
	\end{equation}
In single-shot QEC, a crucial objective is to devise redundant checks with minimal overhead. For this purpose, we consider two methods:
\begin{itemize}
\item[(a)] construct the matrix $A$ directly as a sparse matrix;
\item[(b)] construct  sparse  $\begin{bmatrix}
     H_1 & H_2
 \end{bmatrix} \in \{0,1\}^{\ell'\times (m+\ell)}$ with  $\ell'\ge \ell$ such that
 $ \begin{bmatrix}
     H_1 & H_2
 \end{bmatrix}$ transforms to  $\begin{bmatrix}
     A&{I}_\ell
 \end{bmatrix}$ after Gaussian elimination and row operations, to have an $A$ to define the measurement matrix~\eqref{eq:GB_meas}. 
\end{itemize}

For good BP decoding performance, Gallager suggested using a sparse check matrix with a mean column-weight of at least~3 and a mean row-weight of at least 4~\cite{Gal63}. In line with the idea of two-step decoding~\cite{ALB14,ALB16}, one might initially choose $\begin{bmatrix}
     H_1 & H_2
 \end{bmatrix}$ as an effective classical LDPC matrix  such that $\begin{bmatrix}
     H_1 & H_2
 \end{bmatrix}$ has a mean column-weight of 3.
However,  this approach could introduce an unnecessary surplus of redundant stabilizers.  We will discuss more about the constructions in~Appendix~\ref{app:DS_constructions}.

Our BP decoders employ a unified data-syndrome decoding method, leveraging the entire $\tilde{H}$’s check capabilities and eliminating the need for a two-step decoding process. As a result, 
we propose  constructing a sparse matrix $A$ in method (a)  with a small mean column-weight $\gamma_A=2$ and an adequate mean row-weight $\rho_A$, ensuring a minimal
redundant overhead $\ell/m=\gamma_A/\rho_A$.
By choosing $\gamma_A=2$  we conform to Gallager's recommendation and minimize the number of additional redundant stabilizers in the matrix block $\left[ \substack{ {I}_m\\ A} \right]$. Moreover, the selection of an sufficiently large  value of $\rho_A$ ensures a good classical distance~\cite{MS77,Gal63}.

In the following proposition, we will demonstrate how to construct $A$ as a quasi-cyclic matrix, whose submatrices are  
circulant permutation matrices, derived from random permutations of the identity matrix \cite{Fos04}.
This construction typically leads to an induced Tanner graph with sufficiently large {\it girth} (length of a shortest cycle), which, in turn, results in effective BP performance.
%
\begin{proposition} \label{prop:A}
Consider the $c\times c$ identity matrix $I_c$.  
Let $I_c(p)$ be the circulant permutation matrix constructed from $I_c$ by right-shift each row of $I_c$ by $p$ positions.
Define a binary matrix $A$ of dimension $c\gamma\times c\rho$
as  
\begin{align}
  A=\left[\begin{array}{llll}
    I_c(p_{1,1}) & I_c(p_{1,2}) &\dots &I_c(p_{1,\rho}) \\
    I_c(p_{2,1}) & I_c(p_{2,1}) &\dots &I_c(p_{2,\rho}) \\
    \vdots  & \vdots  &\ddots&\vdots       \\
    I_c(p_{\gamma,1}) & I_c(p_{\gamma,2}) &\dots &I_c(p_{\gamma,\rho}) \\
    \end{array}\right],
\end{align}
where each $0\le p_{i,j}\le c-1$ is an integer. 
Then $A$ is a quasi-cyclic matrix with row-weight $\rho$ and column-weight $\gamma$. 
In particular, $A$ can be specified by a base matrix of dimension $\gamma\times\rho$:
\begin{align}
  \left[\begin{array}{llll}
    p_{1,1} & p_{1,2} &\dots & p_{1,\rho} \\
    p_{2,1} & p_{2,1} &\dots & p_{2,\rho} \\
    \vdots  & \vdots  &\ddots& \vdots       \\
    p_{\gamma,1} & p_{\gamma,2} &\dots &p_{\gamma,\rho} \\
    \end{array}\right].
\end{align}
\qed
\end{proposition}

The parameters $\gamma_A$, $\rho_A$, and $c$ can be chosen so that the number of additional redundant stabilizers    $\ell = c\gamma_A$ is only a small fraction of $m=c\rho_A$.

\subsection{Simulations on generalized bicycle codes}

We consider the $[[126,28,8]]$ GB code in \cite{PK19}
with a $126\times 126$ check matrix. While this matrix already contains 28 redundant rows, it was not specifically designed for single-shot quantum error correction.  Thus we remove 26 of these redundant rows, resulting in a $102\times 126$ quaternary check matrix $H$ over $\{I,X,Y,Z\}$. In accordance with Proposition~\ref{prop:A}, we proceed to construct a quasi-cyclic matrix $A\in\{0,1\}^{34\times 102}$ with parameters $\gamma_A=2$, $\rho_A=6$, and $c=17$. We perform this construction using heuristic random techniques, as in \cite{Mal07}, and successfully create a matrix with a girth of 8, 
 whose base matrix is 
\begin{equation}
    \left[\begin{matrix}
    5 &3 &13 &10 &0 &16\\
    9 &1 &10 &10 &6 &0 \\
    \end{matrix}\right].
\end{equation}
As a result, we obtain a DS check matrix with dimensions of $136\times (126+136)$ in the form of (\ref{eq:S_dec}).
For references, we also provide other DS check matrices in~Appendix~\ref{app:DS_constructions}.

Our error model assumes depolarizing errors with a rate of $\epsilon$ and syndrome bit flip errors with a rate of $\epsilon_\text{b}$. To capture various scenarios, we conduct simulations for the following cases:
\begin{itemize}
\item[(1)] $\epsilon_\text{b} = 0$ (perfect syndrome),
\item[(2)] $\epsilon_\text{b} = \epsilon$  (typical syndrome error rate),
\item[(3)] $\epsilon_\text{b} = 5\epsilon$ to $50\epsilon$ (syndrome noise dominant cases).
     
\end{itemize}
This  is reasonable because, in practical experiments, the measurement error rate can indeed be much worse than the other error gates~\cite{Aru+19,CFYW19}.

\begin{figure}
\centering \includegraphics[width=0.5\textwidth]{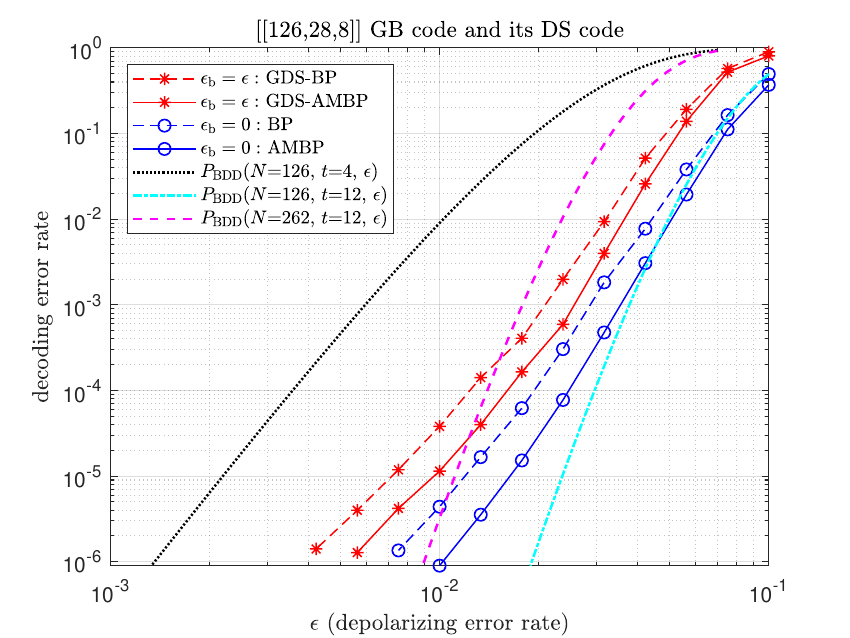} ~~ 
\caption{
    The BP decoding performance of the $[[126,28,8]]$  GB code (with perfect syndromes) and its DS code (when $\epsilon_\text{b} = \epsilon$) using a parallel schedule. 
    The curves for (A)MBP should be compared with the BDD cases for $N=126$. The curves for GDS-(A)MBP should be compared with the BDD case for $N=262$.
} \label{fig:DS_GB}
\end{figure}

\begin{figure}
\centering \includegraphics[width=0.5\textwidth]{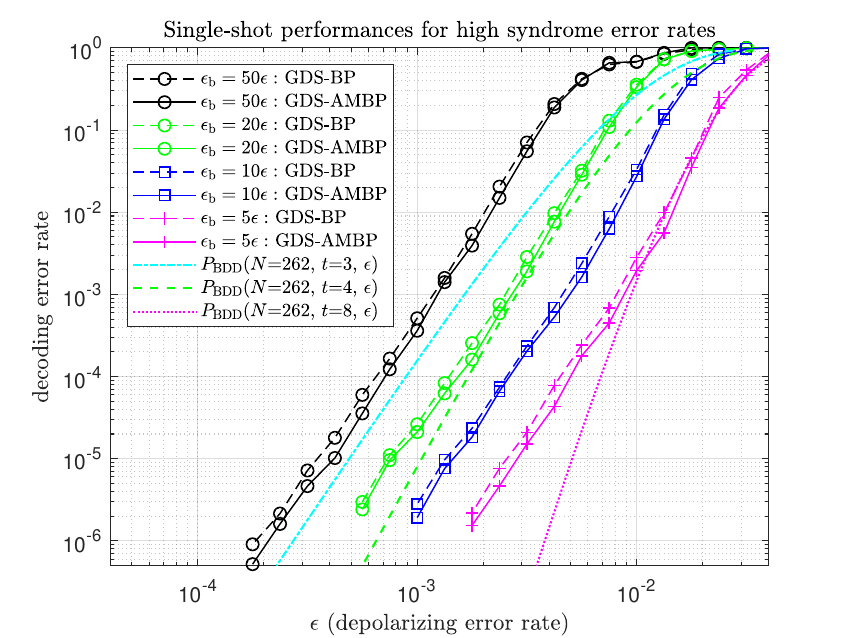} ~~ 
\caption{
    The single-shot performances of GDS  decoders for high syndrome error rates $\epsilon_\text{b} = 5\epsilon$ to $50\epsilon$, using the DS code (with $N=262$ error variables) constructed from the $[[126,28,8]]$  GB code.
} \label{fig:DS_GB_5x--50x}
\end{figure}

We perform simulations for these cases using GDS-MBP with $\alpha=1$ and GDS-AMBP with $\alpha^*\in\{1.4, 1.39, \dots, 0.4\}$. 
We denote GDS-MBP with $\alpha=1$ as GDS-BP, since BP has no added memory effects when $\alpha=1$.
When $\epsilon_\text{b}=0$, we use the original decoders: MBP with $\alpha=1$ (referred to as BP) and AMBP  as in \cite{KL21}.
The maximum number of iterations is set to $T_{\max}=50$. 
This GB code performs effectively with a parallel BP schedule, as shown in \cite{KL21a}, so we adopt the parallel schedule here. Figure~\ref{fig:DS_GB} shows the results for the cases (1)~$\epsilon_\text{b} = 0$ and (2)~$\epsilon_\text{b} = \epsilon$.
Figure~\ref{fig:DS_GB_5x--50x} presents the results for the cases  (3)~$\epsilon_\text{b} = 5\epsilon$ to $50\epsilon$.
For comparison, we also include the bounded distance decoding (BDD) results for different combinations of $N$ and $t$. The BDD decoding error rate  at a certain physical error rate $\epsilon$ is calculated using the formula:	\begin{equation*}  
	P_{\text{BDD}}(N,t,\epsilon) = 1- \sum_{j=0}^t \binom{N}{j} \epsilon^j (1-\epsilon)^{N-j}.
\end{equation*}

Since the distance of the $[[126,28,8]]$ GB code is 8, there are some uncorrectable errors of weight 4.
A reference performance is $P_{\text{BDD}}(N=126,t=4,\epsilon)$ for the perfect-syndrome case.
As can be seen from Figure~\ref{fig:DS_GB},  
BP provides very good performance, while AMBP performs even better, correcting most errors with weight $\le 12$ and nearly achieving the performance of $P_{\text{BDD}}(N=126,t=12,\epsilon)$.
When considering syndrome errors with rate $\epsilon_\text{b}=\epsilon$,   the number of error variables increases to  $N=262$.
 In this scenario, both GDS  decoders are  able to correct most errors with weight 
$\le 12$ and additional high-weight errors, achieving performance that surpasses 
 $P_{\text{BDD}}(N=262,t=12,\epsilon)$ for logical error rates  greater than $10^{-4}$.

When the syndrome noise dominates,
robust performance is crucial, especially the ability to correct most errors of weight up to  $\lfloor\frac{d-1}{2}\rfloor=3$. As shown in Figure~\ref{fig:DS_GB_5x--50x}, when $\epsilon_\text{b} = 5\epsilon$, both GDS  decoders achieve performance close to  $P_{\text{BDD}}(N=262,t=8,\epsilon)$.
Up to $\epsilon_\text{b} = 50\epsilon$, both GDS decoders continue to effectively correct most errors with weight $\le 3$, demonstrating the robustness of the decoder in scenarios with strong syndrome noise during single-shot correction. Since performance is now primarily influenced by the syndrome errors rather than degeneracy, GDS-BP provides competitive performance with lower decoding complexity compared to GDS-AMBP. 
In these syndrome noise cases, the design of a DS matrix is more critical than the choice of decoder. Our proposed design is effective, enabling the use of a low-complexity decoder.

 Note that, for the non-perfect syndrome cases,
since $\epsilon_\text{b}\ge \epsilon$ is assumed, the performance would be dominated by $\epsilon_\text{b}$.
Suppose that $\epsilon_\text{b}=\eta \epsilon$ for  $\eta=1,\,5,\,10,\,20$ and $50$.
Observing the GDS-AMBP curves, we find that, to achieve a decoding error rate below $10^{-5}$, the required physical error rate $\epsilon$ must be lower than  0.01, 0.0028, 0.0014, $7.5\times 10^{-4}$, and $3.2\times10^{-4}$ for $\eta=1,\,5,\,10,\,20$ and $50$, respectively. These values are roughly proportional to $1/\eta$.

When the syndrome error rate is not dominant, such as in the cases of $\epsilon_\text{b}=\epsilon$ or $\epsilon_\text{b}=0$, quantum degeneracy plays a significant role.  
In such scenarios, investing in decoder complexity to fully exploit degeneracy is essential.
An adaptive scheme like GDS-AMBP or AMBP has an improved ability to identify  degenerate errors on the data qubits, leading to enhanced performance by correcting more high-weight errors \cite{KL21}. 
In particular, our decoders can correct most errors up to weight 12, despite that the code only guarantees correction of up to 3 errors.

Finally, we remark that we did not use fixed initialization in the simulations of this subsection. However, based on the current results, it is possible to choose a fixed initialization value like $\epsilon_0 = 10^{-3}$ instead of $\epsilon$ for initialization. This would smooth out some slight curve fluctuations in Figure~\ref{fig:DS_GB_5x--50x}. Using a fixed $\epsilon_0$ for initialization also offers practical convenience by eliminating the need to estimate the physical error rate $\epsilon$.

\section{Conclusion} \label{sec:con}

In this paper, we presented BP algorithms for decoding phenomenological errors with multiple rounds of syndrome extraction, extending the framework of DS codes. 
Our adaptive algorithm GDS-AMBP demonstrated superior performance close to optimal threshold  
for phenomenological quantum memory protected by rotated toric codes compared to MWPM or RG decoders.

It is worth noting that there's potential to reduce the required number of rounds for quantum memory through techniques like a sliding window \cite{DKLP02}. 

In the phenomenological noise model, we made an interesting observation: the assumption of whether the final round of syndrome extraction is affected by data errors becomes less crucial, especially as the number of rounds increases.

We also delved into single-shot quantum error correction and proposed a DS check matrix construction with low overhead. Simulations showed the robust single-shot performance of GDS-(A)MBP on the $[[126,28,8]]$ GB code, which we found to be applicable to many other codes as well.

As a future direction, we aim to generalize the techniques presented in this paper to address the circuit-level noise model, where each location in a quantum circuit might be faulty, and errors propagate through CNOT gates \cite{RHG07,WFSH10}. This ongoing research will extend our understanding of quantum error correction.

It is possible to consider a more generalized phenomenological decoding problem where the syndrome extraction circuit can be flexibly designed for each round, as in Remark~\ref{rmk:H_any}.  If a physical architecture supports this functionality, it will result in  better error correction  performance.
This presents another avenue for further study.

\appendix

\subsection{First iteration of of GDS-MBP}

We show that the calculations in Algorithm~\ref{alg:DS-MBP} with ${\alpha=1}$  provide an approximation to the solution based on the given syndrome.
  We will provide a justification for the first iteration, and subsequent iterations will further improve the approximation as the beliefs are propagated throughout the Tanner graph.

The choice of the normalization parameter $\alpha$ and the fixed inhibition is tailored to achieve good performance with degenerate codes. For a more in-depth discussion on this aspect, please refer to \cite{KL21}.

Consider a GDS check matrix  of the form
    $$\tilde{H} = [H\,|\,H_2],$$ 
where $H\in\{I,X,Y,Z\}^{M'\times N}$ and $H_2\in\{0,1\}^{M'\times M}$.

Assume that we have an error vector $(\bE,\be) = (E_1,\dots,E_N,\, e_1,\dots,e_M) \in \{I,X,Y,Z\}^N\times \{0,1\}^M$, where $E_j$'s and $e_j$'s are independently draw according to
the distributions ${ \{\text{Pr}(E_j=W)\}_{W\in\{I,X,Y,Z\}} }$ for each $E_j$ and ${ \{\text{Pr}(e_j=b)\}_{b\in\{0,1\}} }$ for each $e_j$.

An approximation to the distribution of $(\bE,\be)$, conditioned on a given syndrome vector $\bs\in\{0,1\}^{M'}$, is the product of the conditional marginal distributions of $E_j$'s and $e_j$'s.
Thus we would like to approximately compute the following LLRs
    \begin{align}
    &\Gamma_j^W \approx \ln\frac{\text{Pr}(E_j=I\mid \bs)}{\text{Pr}(E_j=W\mid \bs)}, ~ W\in\{X,Y,Z\}, ~~ j\in\{1,\dots,N\},\\
    &\Gamma_{N+j} \approx \ln\frac{\text{Pr}(e_j=0\mid \bs)}{\text{Pr}(e_j=1\mid \bs)}, ~~ j\in\{1,\dots,M\}.
    \end{align}
We derive the update procedure for obtaining $\ln\frac{\text{Pr}(E_j=I\mid \bs)}{\text{Pr}(E_j=W\mid \bs)}$ in the following. Obtaining $\ln\frac{\text{Pr}(e_j=0\mid \bs)}{\text{Pr}(e_j=1\mid \bs)}$ would be similar.

We have the following initial LLRs according to the channel statistics:
    \begin{align}
    \Lambda_j^W &= \ln\frac{\text{Pr}(E_j=I)}{\text{Pr}(E_j=W)}, ~ W\in\{X,Y,Z\}, ~~ j\in\{1,\dots,N\},\\
    \Lambda_{N+j} &= \ln\frac{\text{Pr}(e_j=0)}{\text{Pr}(e_j=1)}, ~~ j\in\{1,\dots,M\}.
    \end{align}

Recall that we have the sets of neighboring nodes 
$\sN(i) = \{j: \tilde H_{ij}\in\{X,Y,Z\} \text{ or } \tilde H_{ij}=1\}$ for check node~$i$ 
and 
$\sM(j) = \{i: \tilde H_{ij}\in\{X,Y,Z\} \text{ or } \tilde H_{ij}=1\}$ for variable node~$j$.
Note that $\sN(i)$ is the support of row $i$ in $\tilde{H}$.
Let $(\bE,\be)|_{\sN(i)}$ be the restriction of $(\bE,\be)$ to $\sN(i)$, 
and let $\text{$\tilde{H}_i$ denote the $i$-th row of $\tilde{H}$.}$
Then the error syndrome relations $(\bE,\be) * \tilde{H}_i = s_i$ can be 
written as $(\bE,\be)|_{\sN(i)} * {\tilde{H}_i}|_{\sN(i)} = s_i$ for $i=1,\dots,M'$.
For a fixed~$j$, if it participates in check~$i$,
    \begin{align}
    E_j * \tilde{H}_{ij} + (\bE,\be)|_{\sN(i)\setminus j} * {\tilde{H}_i}|_{\sN(i)\setminus j} = s_i, \label{eq:inner_out_j} 
    \end{align}
where $\sN(i)\setminus j$ means $\sN(i)\setminus \{j\}$ to simplify the notation.
For convenience, consider an event $((\bE,\be)=(\bU,\bv))$ for some $(\bU,\bv)\in \{I,X,Y,Z\}^N\times\{0,1\}^M$ that satisfies the syndrome.
Then \eqref{eq:inner_out_j} can be written as 
    \begin{align}
    (\bU,\bv)|_{\sN(i)\setminus j} * {\tilde{H}_i}|_{\sN(i)\setminus j} = I * \tilde{H}_{ij} + s_i, \label{eq:inner_I} 
    \end{align}
  if $U_j=I$,  or 
    \begin{align}
    (\bU,\bv)|_{\sN(i)\setminus j} * {\tilde{H}_i}|_{\sN(i)\setminus j} = W * \tilde{H}_{ij} + s_i, \label{eq:inner_W}
    \end{align}
  if $U_j=W$ for $W\in\{X,Y,Z\}$. These two equations will be used later.

By Bayes rule, 
    \begin{align}
    \ln\frac{\text{Pr}(E_j=I\mid \bs)}{\text{Pr}(E_j=W\mid \bs)} = \ln\frac{\text{Pr}(E_j=I, \text{ syndrome} = \bs)}{\text{Pr}(E_j=W, \text{ syndrome} = \bs)}. \label{eq:LLR_out}
    \end{align}
The term $\text{Pr}(E_j=W, \text{ syndrome} = \bs)$ can be computed as follows:
    \begin{align}
     & \text{Pr}(E_j=W, \text{ syndrome} = \bs) \\
    =& \sum_{(\bU,\bv)\in \{I,X,Y,Z\}^N\times\{0,1\}^M: \atop U_j=W, ~ (\bU,\bv) * \tilde{H} = \bs} \text{Pr}((\bE,\be) = (\bU,\bv)) \\
    =& \sum_{(\bU,\bv)\in \{I,X,Y,Z\}^N\times\{0,1\}^M: \atop U_j=W, ~ (\bU,\bv)|_{\sN(i)} * \tilde{H}_i|_{\sN(i)} = s_i \, \forall i} \text{Pr}((\bE,\be) = (\bU,\bv)). 
    \end{align}
Since 
    $$\text{Pr}((\bE,\be) = (\bU,\bv)) = \prod_{j=1}^N \text{Pr}(E_j=U_j)  \prod_{j=1}^M \text{Pr}(e_j=v_j),$$ 
we have 
  {\EqSize
    \begin{align}
     & \text{Pr}(E_j=W, \text{ syndrome} = \bs) \label{eq:prob_W_s}\\
    \underset{\sim}{\propto}
     & 
     \prod_{i\in\sM(j)} \left( \sum_{ (\bU,\bv)|_{\sN(i)\setminus j}: \atop \text{the condition in \eqref{eq:inner_W}} } \text{Pr}((\bE,\be)|_{\sN(i)\setminus j} = (\bU,\bv)|_{\sN(i)\setminus j}) \right) 
     \notag\\ 
     &
     \times \text{Pr}(E_j=W), \label{eq:apropto}
    \end{align}
  }%
where 
    \begin{align}
     & \text{Pr}((\bE,\be)|_{\sN(i)\setminus j} = (\bU,\bv)|_{\sN(i)\setminus j}) \notag\\
    =& \prod_{j'\in \sN(i)\setminus j: \atop j'\le N} \text{Pr}(E_{j'}=U_{j'})  \prod_{j'\in \sN(i)\setminus j: \atop j'>N} \text{Pr}(e_{j'-N}=v_{j'-N}). \label{eq:prob_out_j}
    \end{align}
The symbol $\left( \underset{\sim}{\propto} \right)$ represents a proportional approximation
and this approximation  is particularly effective when dealing with sparse check matrices.

 A similar approach is applied  {to obtain a form similar to \eqref{eq:apropto}} for approximating  $\text{Pr}(E_j=I, \text{ syndrome} = \bs)$ by using  equation \eqref{eq:inner_I}.

To simplify the notation, we introduce a mixed-alphabet representation of the event:
    \begin{align}
     ((\bE,\be) = (\bU,\bv)) = (\bA = \ba),
    \end{align}
where  $\bA = (A_1,\dots,A_N,\, A_{N+1},\dots,A_{N+M})$ and $\ba = (a_1,\dots,a_N,\, a_{N+1},\dots,a_{N+M})\in \{I,X,Y,Z\}^N\times\{0,1\}^M$.
Then \eqref{eq:prob_out_j} can be written as
    \begin{align}
    &\quad \text{Pr}((\bE,\be)|_{\sN(i)\setminus j} = (\bU,\bv)|_{\sN(i)\setminus j}) \notag\\
    &= \text{Pr}(\bA|_{\sN(i)\setminus j} = \ba|_{\sN(i)\setminus j}) = \prod_{j'\in \sN(i)\setminus j} \text{Pr}(A_{j'}=a_{j'}). \label{eq:prob_out_Aj}
    \end{align}

With $(\bU,\bv)=\ba$, we define two conditions to be used later
    \begin{align}
    \ba|_{\sN(i)\setminus j} * {\tilde{H}_i}|_{\sN(i)\setminus j} = 0, \label{eq:inner_0} \\
    \ba|_{\sN(i)\setminus j} * {\tilde{H}_i}|_{\sN(i)\setminus j} = 1. \label{eq:inner_1} 
    \end{align}

When the Tanner graph is a tree, the proportional approximation in \eqref{eq:apropto} becomes an equality after the beliefs are passed. As mentioned, this approximation typically holds well for sparse check matrices. As a result, we adopt the following update rule by substituting the outcomes  from \mbox{\eqref{eq:prob_W_s}--\eqref{eq:prob_out_Aj}} back into \eqref{eq:LLR_out}:
  {\EqSize
    \begin{align} 
    \Gamma_j^W = \Lambda_j^W + \sum_{i\in\sM(j)} \ln \frac{ 
        \sum\limits_{ (\bU,\bv)|_{\sN(i)\setminus j}: \atop \text{the condition in \eqref{eq:inner_I}} } \left( \prod\limits_{j'\in \sN(i)\setminus j} \text{Pr}(A_{j'}=a_{j'}) \right) 
        }{ 
        \sum\limits_{ (\bU,\bv)|_{\sN(i)\setminus j}: \atop \text{the condition in \eqref{eq:inner_W}} } \left( \prod\limits_{j'\in \sN(i)\setminus j} \text{Pr}(A_{j'}=a_{j'}) \right) 
        }, \label{eq:upd_1}
    \end{align}
  }%
    where the summation term can be simplified as
    \begin{align}
    \sum_{i\in\sM(j): \atop W*\tilde{H}_{ij}=1} (-1)^{s_i} \ln \frac{ 
        \sum\limits_{ \ba|_{\sN(i)\setminus j}: \atop \text{the condition in \eqref{eq:inner_0}} } \left( \prod\limits_{j'\in \sN(i)\setminus j} \text{Pr}(A_{j'}=a_{j'}) \right) 
        }{
        \sum\limits_{ \ba|_{\sN(i)\setminus j}: \atop \text{the condition in \eqref{eq:inner_1}} } \left( \prod\limits_{j'\in \sN(i)\setminus j} \text{Pr}(A_{j'}=a_{j'}) \right) 
        }. \label{eq:upd_2}
    \end{align}

Recall that if the alphabet is quaternary, we compute a scalar message  using $\lambda_W(\lambda^X, \lambda^Y, \lambda^Z)$ as in \eqref{eq:la}.
Notice that
    \begin{align}
    &\lambda_W(\Lambda_j^X, \Lambda_j^Y, \Lambda_j^Z) = \ln\frac{ \sum_{U\in\{I,X,Y,Z\}:U*W=0} \frac{\text{Pr}(E_j=U)}{\text{Pr}(E_j=I)} }{ \sum_{U\in\{I,X,Y,Z\}:U*W=1} \frac{\text{Pr}(E_j=U)}{\text{Pr}(E_j=I)} } \notag\\
    &\quad = \ln\frac{ \sum_{U\in\{I,X,Y,Z\}:U*W=0} \text{Pr}(E_j=U) }{ \sum_{U\in\{I,X,Y,Z\}:U*W=1} \text{Pr}(E_j=U) } \\
    &\quad = \ln\frac{ \text{Pr}(E_j*W=0) }{ \text{Pr}(E_j*W=1) }.
    \end{align}
We extend this function to mixed alphabets: 
    \begin{align}
    \lambda_{(a)}(j) \triangleq \ln\frac{ \text{Pr}(A_j*a=0) }{ \text{Pr}(A_j*a=1) },
    \end{align}
where $a\in\{X,Y,Z\}$ for $j\le N$ and $a=1$ for $j>N$.
Then the update rule mentioned in \eqref{eq:upd_1}--\eqref{eq:upd_2} can be efficiently computed by using the $\boxplus$ operator in \eqref{eq:bsum}:
    \begin{align}
    \Gamma_j^W = \Lambda_j^W + \sum_{i\in\sM(j): \atop W*\tilde{H}_{ij}=1} (-1)^{s_i} \left( \underset{j'\in\sN(i)\setminus j}{\boxplus} \lambda_{(\tilde{H}_{ij'})}(j') \right). \label{eq:upd_3}
    \end{align}
This can be proved by induction. The procedure is straightforward and similar to that in \cite[Appendix~A]{KL21a}. We omit this part for brevity.

For binary variables, $\Gamma_{N+j}$ can be computed similarly as
    \begin{align}
    \Gamma_{N+j} = \Lambda_{N+j} + \sum_{i\in\sM(j)} (-1)^{s_i} \left( \underset{j'\in\sN(i)\setminus j}{\boxplus} \lambda_{(\tilde{H}_{ij'})}(j') \right). \label{eq:upd_4}
    \end{align}

The results in \eqref{eq:upd_3}--\eqref{eq:upd_4} are the same as those in Algorithm~\ref{alg:DS-MBP} for the first iteration. This completes the justification.

\subsection{Constructions of DS check matrices} 
\label{app:DS_constructions}

Herein, we provide additional details on the construction of DS matrices.
Recall that
\begin{itemize}
    \item
    In method (a), a sparse matrix $A$ is constructed and the check matrix 
	$
	\tilde{H} = \small
		\left[\begin{array}{c|cc}
		 {H}   &    {I}_m &    \\
		 &    {A}   &  {I}_\ell \\
		\end{array}\right]
	$
    in \eqref{eq:S_dec} is used  for decoding.

    \item
    In method (b),  we use the  following sparse matrix for decoding
    \begin{equation} \label{eq:H2_dec}
	\tilde{H}' = \small
		\left[\begin{array}{c|cc}
		 {H}   &   I_m &   \\
		      &   H_1 & H_2          \\
		\end{array}\right].
    \end{equation}

 \end{itemize}
 These DS check matrices are different from (\ref{eq:GB_meas}), which corresponds to the stabilizers to be measured.    
 Consequently, the measured error syndrome must be transformed appropriately for decoding.

Using method (a) in Proposition~\ref{prop:A} constructs a matrix $A$ of dimensions $\ell\times m = c\gamma\times c\rho$. 
Similarly, method~(b) can also employ the quasi-cyclic construction  to create the matrix 
$\begin{bmatrix} H_1&H_2 \end{bmatrix}$ 
with dimensions 
$\ell'\times(\ell+m)=c\gamma\times c\rho$.

The $[[126,28,8]]$ GB code is originally defined by 126 checks. We select the first $m$ checks for construction,  ensuring sufficient  rank $(n-k)$ with $m\ge n-k = 98$. 
In both methods, we choose a total of $(m+\ell)$ checks for measurements.

In the following, we have generated several examples by selecting appropriate values for $m$ and adding a suitable number of redundant stabilizers to the 
$[[126,28,8]]$ GB code, resulting in similar values for $m+\ell$.
We aim to construct a check matrix with the largest possible girth. One approach is to slightly extend Proposition~\ref{prop:A}, by incorporating $-1$ in the base matrix to specify a $c \times c$ zero submatrix.

\begin{enumerate}
\item Using  method (a)   with $\ell=34$, $m=102$, and 
$A$ (of dimensions $\ell\times m$)
designed with $(\gamma,\rho,c)=(2,6,17)$, resulting in girth~8.
The base matrix of  $A$  is 
\begin{equation}
    \left[\begin{matrix}
    5 &3 &13 &10 &0 &16\\
    9 &1 &10 &10 &6 &0 \\
    \end{matrix}\right].
\end{equation}

\item Using method (b)   with 
$\ell'=34$,
$\ell=33$, $m=103$, and 
$\begin{bmatrix} H_1&H_2 \end{bmatrix}$ (of dimension $\ell'\times (m+\ell)$) 
designed with $(\gamma,\rho,c)=(2,8,17)$, resulting in girth~8. 
The base matrix of  $\begin{bmatrix}     H_1&H_2 \end{bmatrix}$  is 
\begin{equation}
    \left[\begin{matrix}
    5 &3 &13 &10 &0 &16 &14 &0 \\
    9 &1 &10 &10 &6 &0  &13 &10\\
    \end{matrix}\right].
\end{equation}

\item Using  method (a)   with $\ell=33$, $m=99$, and  $A$ (of dimension $\ell\times m$) designed with $(\gamma,\rho,c)=(3,9,11)$, resulting in girth~6.
The base matrix of  $A$  is 
\begin{equation}
    \left[\begin{matrix}
    3 &0 &7 &3  &9  &7 &5 &6 &4 \\
    5 &8 &6 &10 &10 &5 &5 &1 &7 \\
    2 &6 &0 &0  &1  &8 &1 &0 &2 \\
    \end{matrix}\right].
\end{equation}

\item Using  method (b)  with   $\ell'=33$,
$\ell=33$, $m=99$, and 
$\begin{bmatrix} H_1&H_2 \end{bmatrix}$ (of dimension $\ell'\times (m+\ell)$)
designed with $(\gamma,\rho,c)=(3,12,11)$, resulting in girth~6.
The base matrix of  $\begin{bmatrix}     H_1&H_2 \end{bmatrix}$  is 
\begin{equation} \setcounter{MaxMatrixCols}{20}
    \left[\begin{matrix}
    3 &0 &7 &3  &9  &7 &5 &6 &4 &0  &3  &9 \\
    5 &8 &6 &10 &10 &5 &5 &1 &7 &5  &7  &-1\\
    2 &6 &0 &0  &1  &8 &1 &0 &2 &-1 &-1 &9 \\
    \end{matrix}\right].
\end{equation}

\end{enumerate} 
 In our tests, the first case, which is our chosen approach, demonstrated the best performance. 
These results  further support our recommendation of method (a)   to construct a quasi-cyclic matrix  $A$ with a mean column-weight of $\gamma_A=2$.

Note that  there are predefined row operations needed to transform the 
DS check matrices for measurements and for decoding.
For example, in the above case 2), $\begin{bmatrix}
    H_1&H_2
\end{bmatrix}$ has 34 rows but is of rank 33. We convert it to a systematic form $\begin{bmatrix}
    A&I
\end{bmatrix}$ with 33 rows and use this $A$ to define the DS measurements in~\eqref{eq:GB_meas}.  In this case, there are 103+33 stabilizers for measurements as in~\eqref{eq:GB_meas}, while 103+34 checks will then be generated for decoding, as in \eqref{eq:H2_dec}. These two matrices, \eqref{eq:GB_meas} and \eqref{eq:H2_dec}, transform into each other using the predefined row operations, which can convert a measured syndrome to a syndrome for decoding.


\begin{IEEEbiographynophoto}{Kao-Yueh Kuo}
	(Member, IEEE) originally from Hsinchu, Taiwan, completed his Ph.D. in Electrical Engineering at National Tsing Hua University in 2015. He has extensive experience in communication chip design industry for more than 12 years, before and after his Ph.D. study. Then he pursued postdoctoral research at National Yang Ming Chiao Tung University starting in 2019. In 2024, Kuo joined the University of Sheffield’s School of Mathematical and Physical Sciences in the UK as a research associate. His research is primarily focused on classical and quantum communications, with a special emphasis on coding theory and iterative decoding methodologies.
\end{IEEEbiographynophoto}

\begin{IEEEbiographynophoto}{Ching-Yi Lai}
	(Senior Member, IEEE)  received his M.S. in 2006 and B.S. in 2004 from National Tsing Hua University, Taiwan and completed his Ph.D. at the University of Southern California in 2013. Following   postdoctoral   positions at University of Technology Sydney and Academia Sinica, he joined the Institute of Communications Engineering at National Yang Ming Chiao Tung University, Hsinchu, Taiwan. In 2022, Dr. Lai was promoted to Associate Professor. Dr. Lai was honored with the Young Scholar Fellowship from the National Science and Technology Council, Taiwan. His research explores diverse areas, including quantum coding theory, quantum information theory, fault-tolerant quantum computation, and quantum cryptography.
\end{IEEEbiographynophoto}

\end{document}